\documentclass[twocolumn]{aastex6}

\epsscale{1.17} % Set figure width

\usepackage{epsfig}
\usepackage{amsmath}
\usepackage{amssymb}

\newcommand{\Msolar}{\mbox{$\rm M_{\odot}$}} % solar mass
\newcommand{\Rsolar}{\rm{R_{\odot}}} % solar radius
\newcommand{\Lsolar}{\rm{L_{\odot}}} % solar luminosity
\newcommand{\iso}[2]{\mbox{$^{#1}{\rm #2}$}} % isotope designation

\shorttitle{HeWD+RGB mergers}
\shortauthors{Zhang et al.}

\begin{document}
\title{Population synthesis of helium white dwarf--red giant star mergers and the formation of lithium-rich giants and carbon stars}

\author{Xianfei Zhang}
\affil{Department of Astronomy, Beijing Normal University, Beijing, 100875, China}
\email{zxf@bnu.edu.cn}

\author{C.~Simon Jeffery}
\affil{Armagh Observatory, College Hill, Armagh BT61 9DG, UK;\\
School of Physics, Trinity College Dublin, Dublin 2, Ireland}

\author{Yaguang Li}
\affil{Department of Astronomy, Beijing Normal University, Beijing, 100875, China}

\and

\author{Shaolan Bi}
\affil{Department of Astronomy, Beijing Normal University, Beijing, 100875, China}

%\altaffiltext{}{}

\begin{abstract}
The formation histories of lithium-rich and carbon-rich red giants are not yet understood.
It has been proposed that the merger of a helium-core white dwarf with a red giant branch star might provide a solution.
We have computed an extended grid of post-merger evolution models and combined these with predictions of binary star population synthesis.
The results strongly support the proposal that the merger of a helium white dwarf with a red giant branch star
can provide the progenitors of both lithium-rich red clump stars and early-R carbon stars.
The distribution of post-merger models in $T_{\rm eff}$, $\log g$, $\log L$,  the surface abundances of lithium and carbon, and the predicted space densities agree well with the observed distributions of these parameters for Li-rich and early-R stars in the Galaxy.
\end{abstract}

\keywords{stars: binary:close---stars: abundances---stars: chemically peculiar---star:evolution---white dwarfs}

\section{Introduction}

%\noindent {\bf Key to annotations}\\
%{\it CSJ - text in italics are my suggestions or questions }\\
%{\bf text in bold are changes which need to be checked or completed (eg references}\\
%{\red text in red is original text  I didn't understand and have not yet fixed.}

Big Bang nucleosynthesis produced most of the observed lithium in the universe.
Lithium is destroyed in stellar interiors once the temperature exceeds $2.5\times10^6 \rm K$.
Both theory \citep{Iben67b,Iben67a} and observations of globular cluster stars \citep{Lind2009} show that
surface lithium will be depleted during red giant branch (RGB) evolution to less than the solar surface value of $\rm A(Li)\equiv\log_{10}(N(Li)/N(H))+12=1.5$
due to the deep convective envelope in the first dredge-up stage.
However, a small fraction of red giant branch stars show a lithium-rich (Li-rich) photosphere ($\le 1\%$ in the Galactic disk, \citealt{Brown89}).
Hundreds of Li-rich giants have been found in the last few years, including tens of super Li-rich stars with $\rm A(Li)\ge 3.2$
\citep{Kumar2011,Adamow2014,Casey2016,Deepak2019,Zhou2019,Singh2019b}.
The most Li-rich giant star so far discovered has $\rm A(Li)=4.5$ \citep{Yan2018}.
Most such super Li-rich giants are identified as red clump stars.
Since red clump stars are core-helium burning stars, they should have already passed through the first red-giant branch
where most of their surface lithium would have been destroyed \citep{Jofre2015,Carlberg2015,Kumar2018,Singh2019a,Singh2019b}.
As well as Li-rich giants observed as field stars, some have been found in clusters \citep{Carlberg2015},
and in metal-poor populations \citep{Li2018}.
Thus, the formation of Li-rich giants does not depend on special initial conditions but remains
a puzzle for stellar evolution theory.

Three principal scenarios have been proposed to explain the lithium enrichment in these cases.
(1) The lithium has been gained from a brown dwarf or rocky planet companion which contains lithium \citep{Ashwell2005,Aguilera2016}.
(2) Lithium-rich material has been accreted from an asymptotic-giant branch (AGB) star or a nova companion  \citep{Jose1998}.
(3) The lithium has been produced in the interior of the star and lifted to the surface by enhanced extra-mixing
\citep{Cameron71,Sackmann1999,Charbonnel2000,Denissenkov2004,Yan2018}.
In scenarios (1) and (2), we may expect to observe some other elements from their companion.
The typical upper limit for lithium enrichment by such means is $\rm A(Li)\le 2.2$ \citep{Aguilera2016}.
In scenario (3), the fresh lithium is produced by the Cameron-Fowler mechanism \citep{Cameron71},
in which lithium is created following the ${\rm ^{3} He}(\alpha,\gamma){\rm ^{7}Be}$ reaction;
beryllium  is assumed to be transported to a cooler region where it is converted to lithium by $\beta$-decay
(${\rm ^{7}Be}(e^{-},\nu){\rm ^{7}Li}$).
Thus, such a star should have a hot burning interior zone  to create  ${\rm ^{7}Be}$ and an extra-mixing convection zone
to bring the ashes of nuclear activity to the surface while avoiding lithium destruction by proton capture.
In this scenario (3), the actual mechanism that drives the extra mixing plays an important role.
Several have been proposed including rotation and internal instabilities.
So far, details are lacking and the frequency at which such mechanisms occur in red giants is not known.

\begin{figure*}
\centering
\includegraphics [angle=0,scale=0.5]{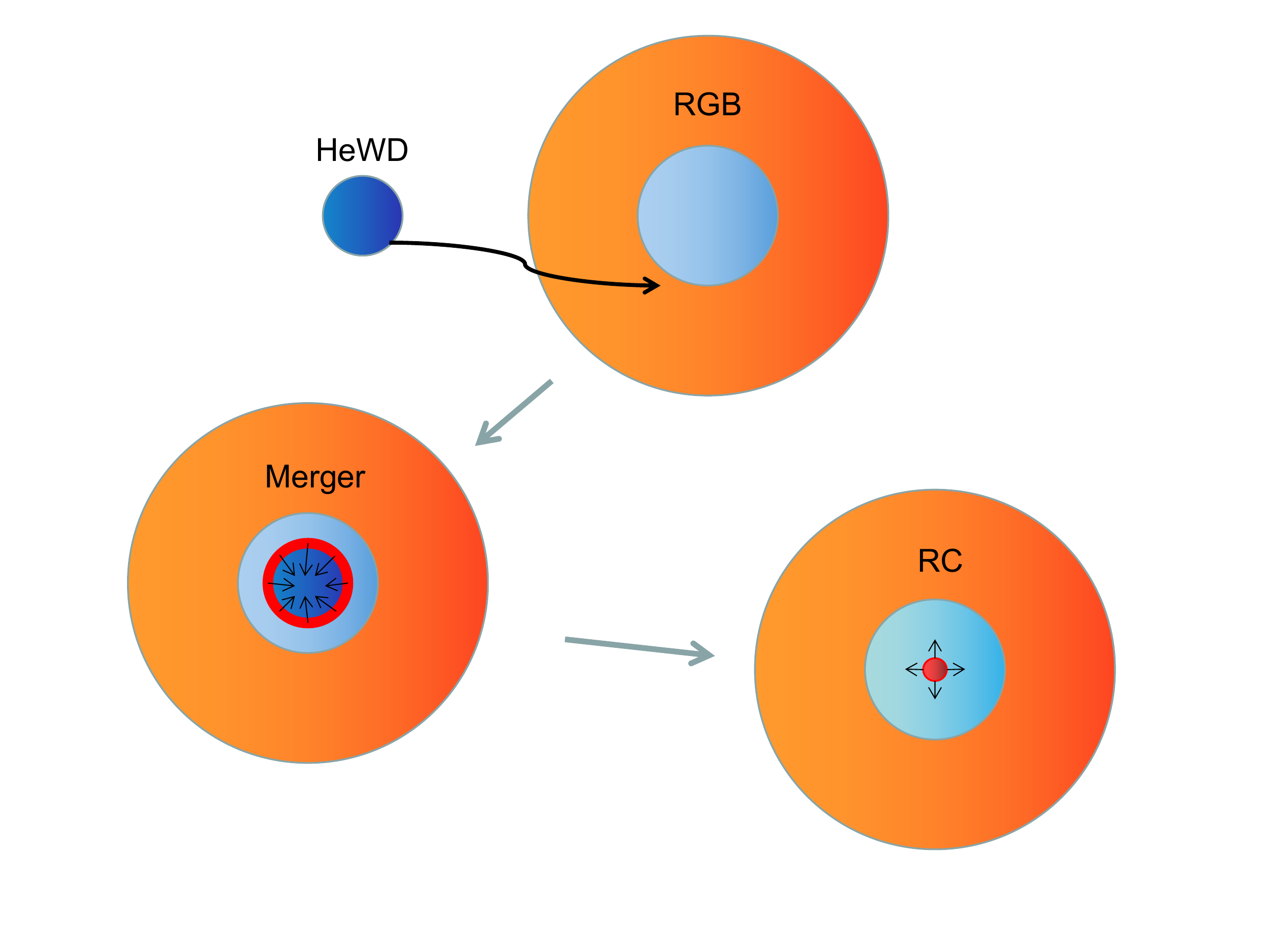}
\caption{Schematic representation of possible steps in a helium white dwarf (HeWD) plus red-giant branch (RGB) star merger
leading to the formation of a red clump star with a lithium-rich or carbon-rich surface. Orange represents the hydrogen-rich envelope, blue is helium-rich material, and red represents the nuclear helium-burning region.}
\label{1}
\end{figure*}

Another type of star with chemically peculiar surface are the carbon stars. Carbon stars are normally classified as
being one of spectral types N, R, and J, \citep{Secchi1868,Fleming1908} and have $\rm C/O>1$ by numbers in their atmospheres.
The N-type and some cool R-type (late-R) carbon stars are recognized as the normal and most significant population of carbon stars.
Their surface compositions can be reproduced by low-mass AGB nucleosynthesis models which have carbon enriched by third dredge-up
and have s-process elements in their atmospheres \citep{Zamora2009}.
The hot R-type stars (early-R stars) and J-type stars are different from normal carbon stars.
Their surfaces are enriched in $\rm ^{13}C$($\rm ^{12}C/^{13}C<15$) as well as  \iso{12}{C}
but without s-process enhancements \citep{Dominy1984,Zamora2009}.
The early-R stars have magnitudes similar to the red clump stars \citep{Knapp2001}.
Most of the early-R stars are single stars.
The J-type carbon stars show a high luminosity with a location close to the AGB on the HR-diagram, and have a smaller ratio
of $\rm ^{12}C/^{13}C$ and more lithium on the surface than early-R stars \citep{Abia2000}.

In the standard stellar evolution theory it is difficult to obtain models of single stars with carbon enriched surfaces
unless they are in the thermally pulsing AGB stage or in the Wolf-Rayet stage of massive star evolution.
The R-stars are likely in the red clump (core helium-burning) phase and are too faint to be in either of those stages.
The formation channel of early-R stars was unclear for a long time, until the proposal that they formed by binary-star mergers
was supported by binary star population synthesis calculations \citep{Izzard2007}.
A further study with full calculation of the post-merger evolution with details of stellar parameters and abundance further supported
this channel \citep{Zhang2013}.
In this merger model, the early-R stars are formed by helium white dwarfs (HeWD) which merge with a red giant branch (RGB) star.
Surface lithium is enriched during this merger process.
This channel could also produce the single J-type carbon stars \citep{Zhang2013},
though these stars could also be explained by pollution from a nova companion \citep{Sengupta2013}.
Due to the character of Li-rich giant and early-R stars,
i.e. red giant-like stars with surfaces enriched in lithium and/or carbon with a low  $\rm ^{12}C/^{13}C$ ratio,
it seems likely that the HeWD+RGB merger channel can produce either or both of these of stellar types.

\citet{Piersanti2010} performed a three-dimensional smoothed particle hydrodynamics (3D SPH)
simulation of a low-mass HeWD+RGB merger ($M_{\rm HeWD}\le 0.2 \Msolar$)
which shows that no efficient helium burning occurred to dredge up carbon material to the surface.
\citet{Zhang2013} found similar results from their 1D post-merger calculation and also suggested that
a higher mass helium white dwarf subducted into a low-core-mass red
giant could produce carbon enrichment at the surface of the giant.
However, only five models were investigated by \citet{Zhang2013},
which also adopted a much higher accretion rate than that indicated by the 3D SPH simulation.

To survey the products of HeWD+RGB mergers in a larger parameter space, we extended the study of \citet{Zhang2013} by
 modifying the model of the merger process and calculating models for a wider range of progenitor binaries.
We have adopted the results of the SPH simulation into the 1D stellar evolution calculation
and combined the resulting evolutionary tracks with the results of binary star population synthesis.
We aim to compute the statistics and surface abundance of the post-merger systems,
and compare these with  observational data for Li-rich giants and early-R carbon stars.
We aim to answer: when and how does a merger make lithium or carbon?
how does the fresh material get to the surface?
and how many of Li-rich / early-R stars should be  observed in the Galaxy?

In this paper, we define a Li-rich giant star
as having $\rm A(Li)=\log_{10}(N(Li)/N(H))+12>1.5$
and a carbon star as having $\rm N(C)/N(O)>1$
by numbers. In \S\,\ref{s_models}, we introduce the methods of constructing and
evolving the models of post-mergers.
The comparison of theory with observation of Li-rich giants and early-R stars
is shown in section \S\,\ref{s_results}. The discussion and conclusions
are in \S\,\ref{s_conclusion}.

\section{Modelling the mergers}
\label{s_models}

Fig.~\ref{1} shows a schematic sequence of events during a HeWD+RGB merger.
Once the HeWD comes into contact with the expanding red giant, a common envelope forms.
The HeWD will merge with the helium core of the giant if spiral-in occurs before the entire envelope is ejected.
Then a single giant-like star forms  which contains a hybrid-core surrounded by an hydrogen-rich mantle.
The structure of hybrid-core should be similar to the product of a double HeWDs merger,
i.e., a degenerate core originating from the HeWD,
surrounded by a very hot shell ($\rm >10^8K$) from the disrupted helium core of the progenitor red giant.
Subsequently, the core will be heated by a series of He-burning shell flashes which burn inwards towards the centre.
Finally, a red clump star forms once helium core burning has been established.

To represent such an evolutionary path in detail,
three key steps are proposed in our model calculation.
In step 1, information about potential merger progenitors is obtained from binary star population synthesis (BSPS).
In step 2, a grid of evolutionary tracks and surface abundances is calculated for an ensemble of post-mergers models.
In step 3, the characteristic properties of post-merger stars are calculated by
combining the results of BSPS and the grid of post-merger evolutionary tracks.

\subsection{Step 1: binary population synthesis}

To obtain details of the HeWD+RGB merger channel, we adopt BSPS to investigate
whether the rate of possible mergers is sufficiently high to make
a significant contribution to the population of Li-rich and C-rich giants.
At same time, we calculate the masses of the HeWD and RGB progenitors which will be
used for the subsequent calculation of post-merger evolution (Step 2).

We model populations of binary stellar stars with the
\texttt{BSE} code \citep{Hurley00,Hurley02},
starting from zero-age main-sequence stars.
The basic parameters for binary star evolution in the rapid evolution code in this work
are chosen to be the same as those previously used to model
the R stars by \citet{Izzard2007}, which are also similar to those used by \citet{Politano2008,Politano2010}.
We evolve $10^7$ pairs of stars within an evolution time of $14\,\rm{Gyr}$
and record the properties of HeWD+RGB binaries at the onset of the common envelope (CE) phase.
The mass distribution of such pre-CE binaries will be
used to set the grid of parameters for the calculations in steps 2 and 3.

We generate the mass ratio distribution of zero-age binary stars according to the formula of \citet{Eggleton89}
and adopt the initial mass function of \citet{Miller79} in the range $0.08-100\,\Msolar$.
The distribution of orbital separations, $p(a)$, is calculated by the formula of \citet{Han98}:
$$ p(a)=\left\{
\begin{array}{lcl}
0.070(a/a_0)^{1.2}      &      & {a \leq a_0}\\
0.070                   &      & {a_0 \le a \le a_1},\\
\end{array} \right. $$
where $a_0=10\,\Rsolar$, $a_1=5.75 \times 10^6\,\Rsolar = 0.13\,\rm{pc}$.
The parameters chosen in this work have been used in several previous studies
on double WD mergers in the Galaxy \citep{Han98, Zhang2014};
hot subdwarfs \citep{Han02,Han03,Zhang12a};
type Ia supernova \citep{Wang2009} and EL CVn stars \citep{Chen2017}.

\begin{figure}
\includegraphics [angle=0,scale=0.5]{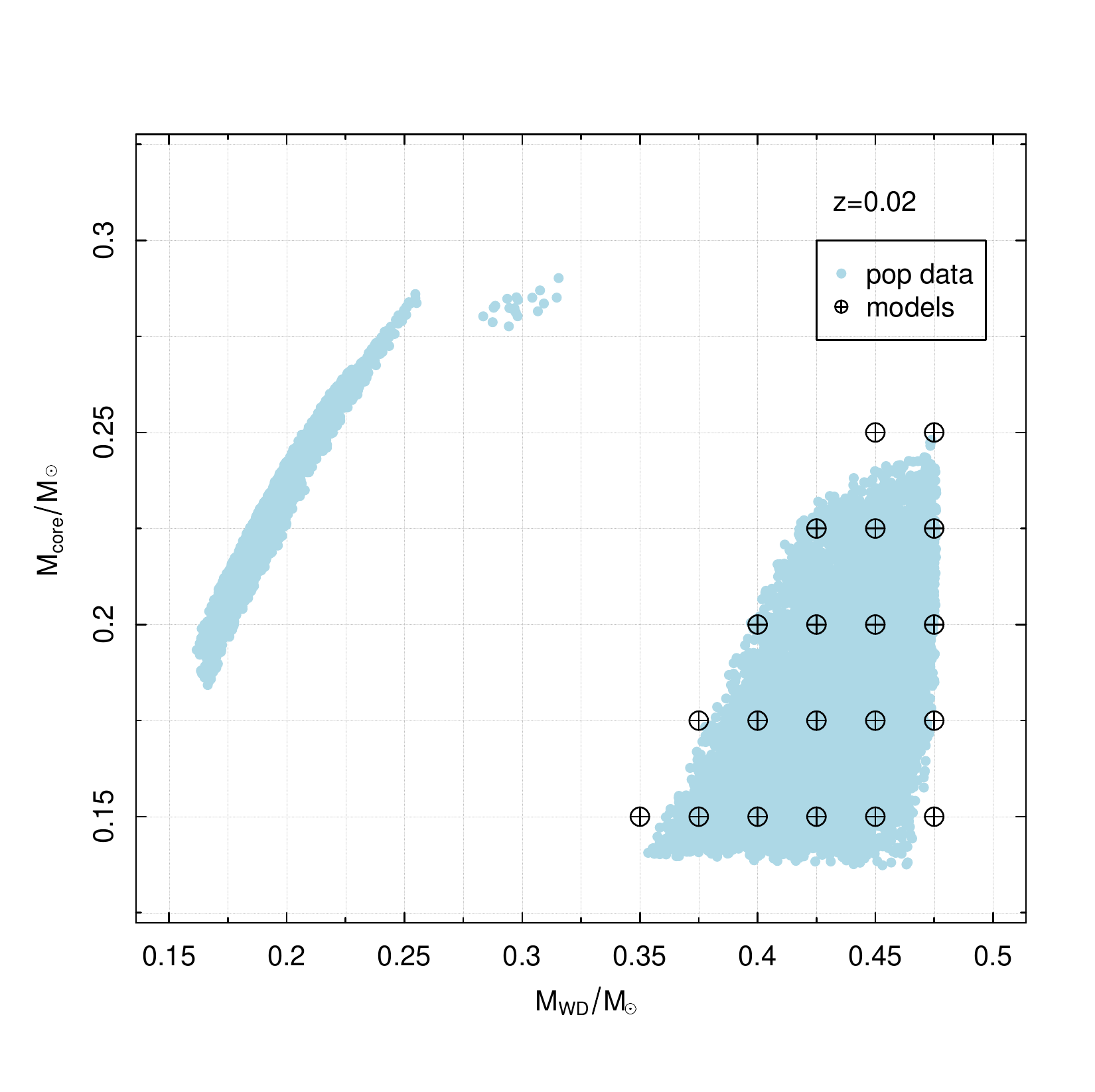}
\caption{Models of HeWD+RGB merger remnants in the $\rm M_{WD}+M_{core}$ plane.
The circles with crosses indicate the merger models in this work.
The dots indicate the masses of HeWD and core masses of the RGB stars
in the pre-common envelope phase obtained in the binary population synthesis.
The metallicity is $Z=0.02$.}
\label{2}
\end{figure}

\begin{figure}
\includegraphics [angle=0,scale=0.5]{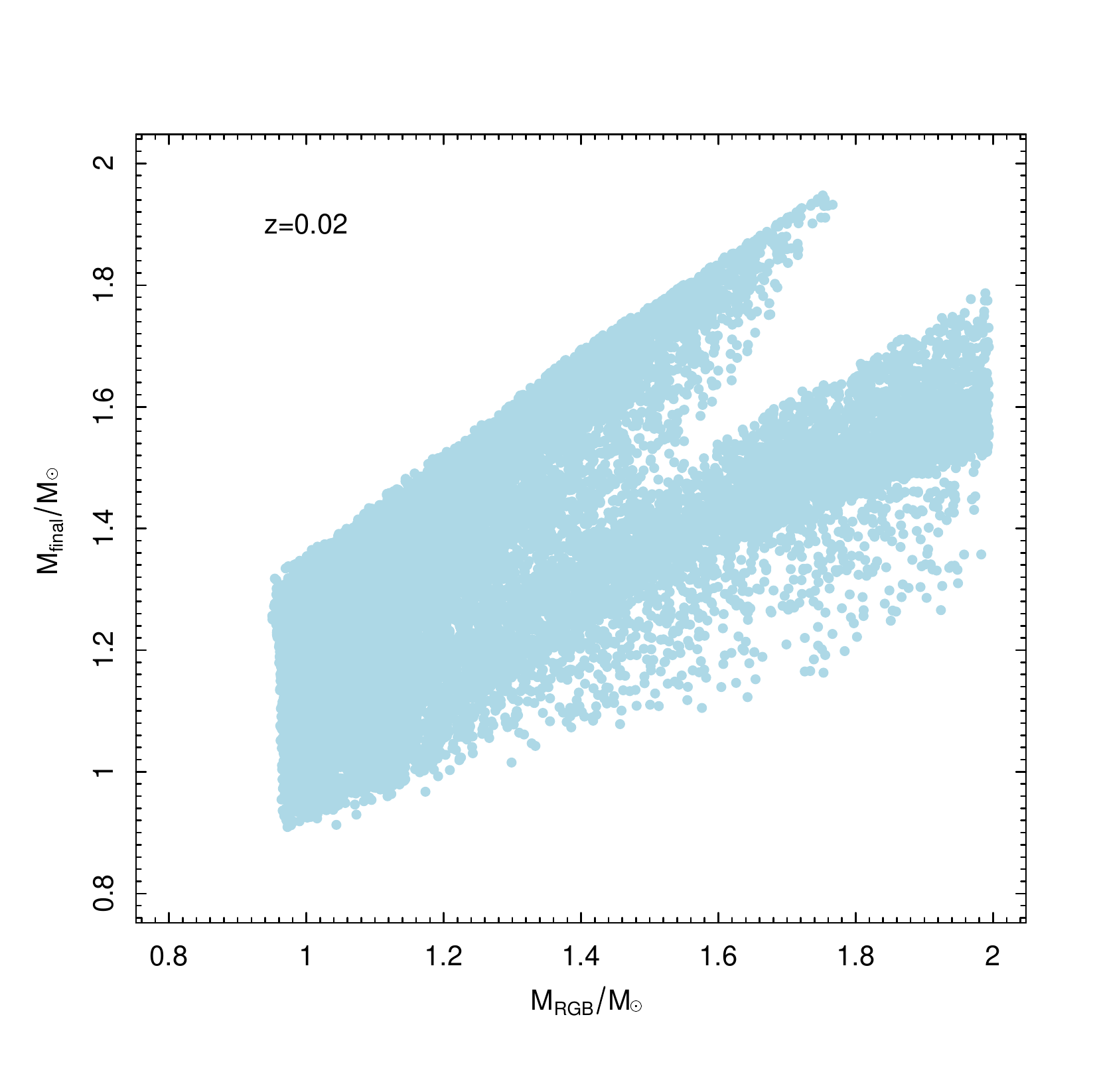}
\caption{The initial and final masses of RGB stars in the pre-common envelope and
post-merger phases. The metallicity is $Z=0.02$. }
\label{3}
\end{figure}

Fig.~\ref{2} shows the mass distribution of HeWDs and RGB cores with a metallicity of $Z=0.02$ in the
pre-common envelope phase.
This is comparable with the results of \citet{Izzard2007}.
As in \citet{Izzard2007}, the gap between the two populated regions is related
to the initial periods of progenitor main-sequence binaries.
The low-mass HeWDs form from short period (few days)
binaries and the high-mass HeWDs form from long period (hundreds of days)
binaries.
The masses of the RGB star progenitors lie in the range 1--2\Msolar, and will reduce after common envelope ejection.
The final mass of the mergers shown in Fig.~\ref{3} lies in the range 0.9--2\Msolar.

\subsection{Step 2: post-merger evolution}

To examine features of the post-merger stars including the enrichment of elements, we
used the stellar evolution code {\sc mesa} version V8118
\citep[Modules for Experiments in Stellar Astrophysics;][]{Paxton11, Paxton13, Paxton15}.
To obtain the initial HeWDs, we evolved a 1.2\Msolar zero-age MS star until the He core reaches one
of 0.350, 0.375, 0.400, 0.425, 0.450 or 0.475\Msolar.
Nucleosynthesis is switched off and a high
mass-loss rate is applied to remove the hydrogen envelope
completely, leaving a model of an exposed He core.
Then the remnant core is allowed to cool to become a white dwarf,
finally ending with a model having a surface luminosity $\log (L/\Lsolar) = -2$ \citep{Zhang12a}.
According to \citet{Zhang2013}, surface enrichment following a  merger only occurs for high-mass HeWDs.
We also did a test for low-mass HeWD merger model and confirmed such results.
Hence, we have here chosen the masses of HeWDs to lie in the range
$0.350 - 0.475 \Msolar$ in steps of 0.025$\Msolar$.

It is challenging to simulate a merger with a 1D stellar evolution code,
however a series of separate accretion steps is proposed to represent such
a merger process and has previously been used successfully to represent some observations of merger remnants
 \citep{Zhang12a,Zhang2013,Zhang2014,Zhang2017}.
We separate the merger process into two accretion steps in this work, i.e.,
(1) accretion of helium-rich materials {\it onto the HeWD} to represent the process of the HeWD+RGB-core merger in the deep interior
and then
(2) accretion of hydrogen-rich materials to represent the assimilation of the surrounding RGB envelope by the new hot hybrid core.
The fractional distribution of elements within the helium-rich material is computed from the average mass fractions of a 0.2\Msolar HeWD.
The equivalent distribution within the hydrogen-rich material is computed  from the average mass fraction of an envelope of a 1.2\Msolar RGB star.
The accretion rate during the helium accretion phase is $10^{-3} \Msolar \rm yr^{-1}$
which is similar to the average accretion rate obtained in the SPH simulation \citep{Piersanti2010}.
An accretion rate of $1 \Msolar \rm yr^{-1}$ is used to re-establish the hydrogen-rich envelope around the merged core, as was used by \citet{Zhang2017}.
In all, the whole process takes place on a similar time-scale to a CE phase \citep{Ivanova2011,Passy12,Ivanova2013}.
The masses of the RGB helium cores range from 0.150 to 0.25 $\Msolar$ in steps of 0.05$\Msolar$, as shown in Fig.~\ref{2}. In general, we set the initial hybrid-core masses of mergers to be the sum of masses of HeWD and RGB core, which will be modified by envelope convection during the merger process. The final masses of the merger combines contributions from the hybrid-core and the envelope masses.
The masses of final merged models range from 0.80 to 2.0 $\Msolar$ in steps of 0.1$\Msolar$ to cover the distribution of the final masses from binary population synthesis of step 1 (see Fig.~\ref{3}).
Thus, we obtained 259 initial models of post-merger stars
for each metallicity, i.e., Z=0.03, 0.02, 0.01 and 0.004.

For the subsequent post-merger evolution, we adopted parameters
similar to the MESA isochrones and stellar tracks (MIST) project
for normal stars \citep{Dotter2016,Choi2016} (see Appendix A).
We set the mixing length parameter $\alpha = l/H_{\rm p} = 1.82$.
We adopt the OPAL Type 2 opacity tables in order to account for the changing abundances
of carbon and oxygen following the He flashes\citep{Iglesias1996,Ferguson2005}.
We chose the {\it simple photosphere} option for the outer boundary condition (i.e. $T^4(\tau)=3/4\,T_{\rm eff}^4(\tau+2/3)$).
In our models, mixing is by convection in the convective regions and atomic diffusion
in the radiative areas\citep{thoul94}. Diffusion includes the processes of
gravitational settling, thermal diffusion, and concentration diffusion.
The atomic diffusion coefficients are those calculated by \citet{paquette86}.
We also consider semiconvective and thermohaline mixing as in MIST.
The mass loss rate of the post-merger stars is computed according to Reimers' formula with $\eta_{\rm R} = 0.5$.
Nuclear reaction networks follow the abundances of $21$ species: \iso{1}{H}, \iso{2}{H}, \iso{3}{He}, \iso{4}{He}, \iso{7}{Li}, \iso{7}{Be}, \iso{8}{B}, \iso{12}{C}, \iso{13}{C},
\iso{13}{N}, \iso{14}{N}, \iso{15}{N}, \iso{16}{O}, \iso{17}{O}, \iso{18}{O}, \iso{19}{F}, \iso{22}{Ne}, \iso{23}{Na},
\iso{24}{Mg}, \iso{27}{Al} and \iso{56}{Fe}.

\subsection{Step 3: population synthesis of post-mergers}

We have the masses of progenitors from step 1 and a grid of evolutionary tracks of post-mergers from step 2.
Hence, we can calculate the distribution of the post-merger models by combining both steps, from which
we can obtain all the required parameters, i.e., the masses, surface effective temperature,
surface gravity, and surface abundances.
As shown in Fig.~\ref{3} the final masses of the merged stars lie in a relatively small range.
Thus the evolutionary tracks are very close to each other for red giant evolution.
But the variation of surface abundance is much more complex.

Therefore, we do not interpolate the paths for different masses,
but instead we assume that stars with similar HeWDs, core masses, and final masses
are on one of the theoretical evolutionary tracks that we have computed.
We then calculate the number of merger remnants on each track from the BSPS prediction.
We determine the evolutionary stage of each BSPS star along each track by a
random sample selection which includes a probability weighted by the timescale associated with each point on the track.

%------------------------------------------------------------------------------

\section{Results}
\label{s_results}

\begin{figure}
\includegraphics [angle=0,scale=0.5]{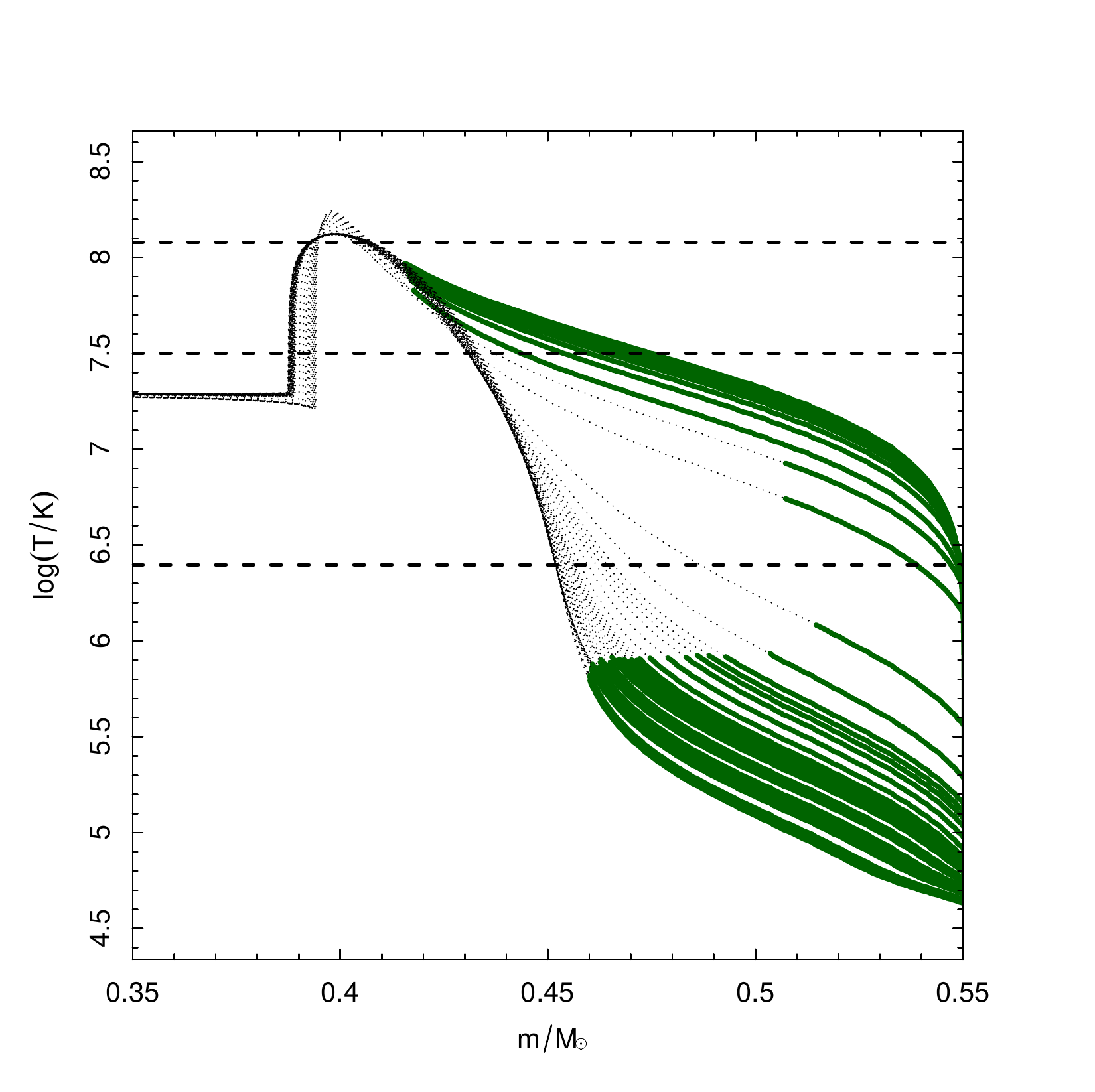}
\caption{The temperature profile of a 0.40\Msolar HeWD + 0.15\Msolar\ helium-core RGB model,
during the hydrogen-{\it assimilation} process. The dotted line shows the first 50 profiles of the model;
the time increases from top to bottom. The thick (green) line indicates the convection dominated region in each profile.
Three dashed lines indicate the {\it approximate minimum} temperatures for the $3\alpha$ reaction ($T=1.2\times 10^8 K$, e.g., \citet{Suda2011}),
Cameron-Fowler mechanism reaction ($T=3\times 10^7 K$, e.g., \citet{Sackmann1992})
and lithium destruction ($T=2.5\times 10^6 K$, e.g.,\citet{Pinsonneault1997}), respectively.
During the interval shown here, the mass of the hydrogen-envelope (not shown) increases from 0 to 0.02 \Msolar.
}
\label{4}
\end{figure}

\begin{figure}
\includegraphics [angle=0,scale=0.5]{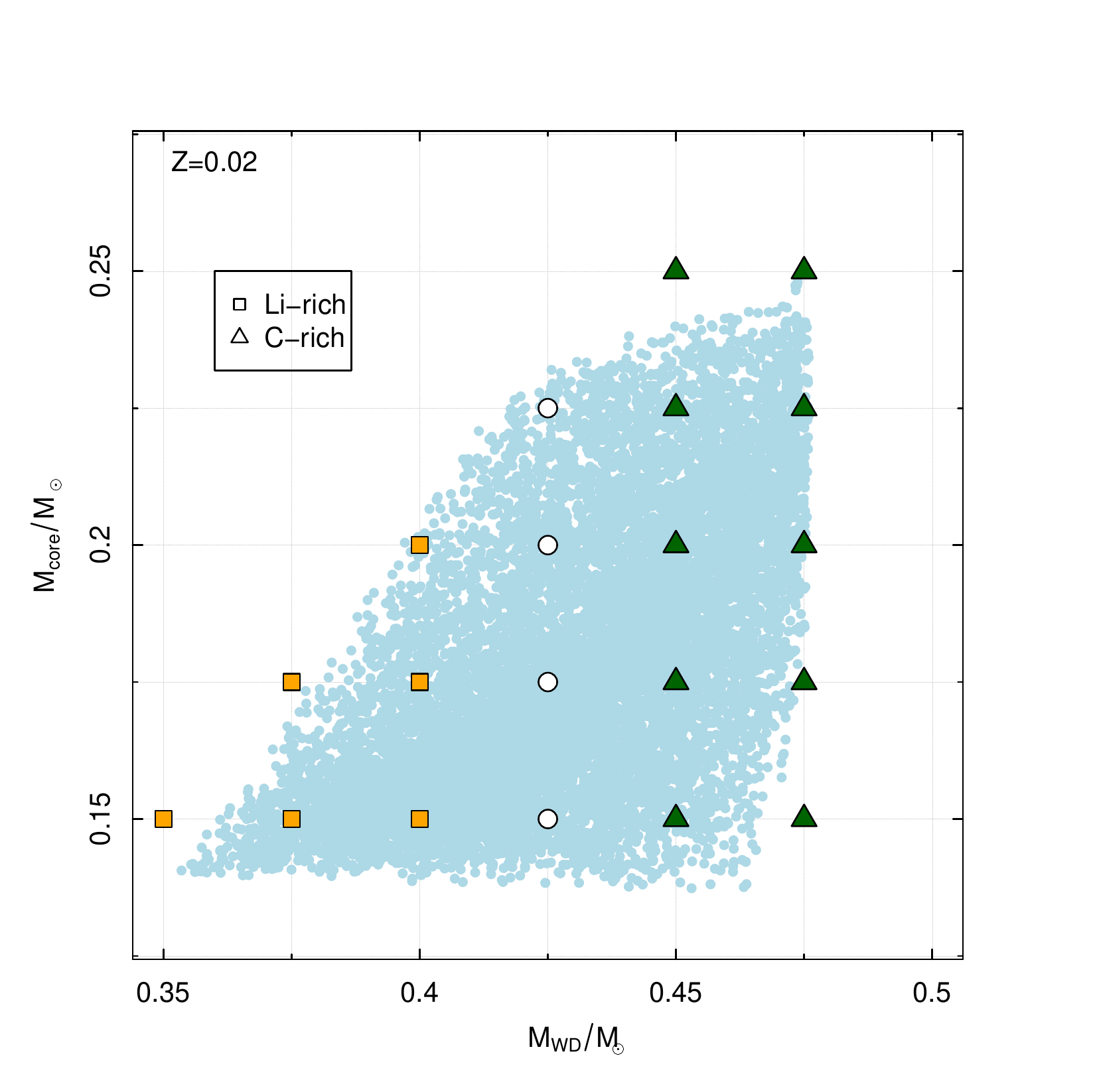}
\caption{Similar to Fig.~\ref{2}, models of merged HeWD+RGB  remnants in the $\rm M_{WD}-M_{core}$ plane.
The squares (orange) indicate Li-rich merged models and triangles (green) show C-rich merged models.
The circles indicate unenriched models.
The metallicity is Z=0.02.}
\label{5}
\end{figure}

According to the common-envelope merging process,
the remnant contains a hybrid core with a hot helium shell ($>10^8\rm K$)
surrounded by a hydrogen envelope.
We adopt two separate fast-accretion phases to compute the structure of such a post-merger model.
The surface abundance is enriched during the second (hydrogen) accretion phase.

Fig.~\ref{4} shows the temperature structure of the hybrid core during the second accretion phase
in the 0.40\Msolar HeWD + 0.15\Msolar\ helium core RGB model.
During the first $\sim 10$ steps of this phase, the convection zone can extend from the merged-star surface down to the hot helium shell.
Hence some material in the envelope can reach the hot layer (where the temperature $T>10^7\rm K$) and hence participate in nuclear reactions during this short interval.
Hence, \iso{3}{He} from the hydrogen envelope is mixed with \iso{4}{He} in
hot helium shell and produces the fresh \iso{7}{Li} by the
${\rm ^{3}He}(\alpha,\gamma){\rm ^{7}Be}(e^{-},\nu){\rm ^{7}Li}$ reaction
{\bf once  $T>3\times 10^7\rm K$}.
Meanwhile, fresh \iso{7}{Li} will be destroyed by the
${\rm ^{7}Li}(p,\alpha){\rm ^{4}He}$ reaction regions where $T>2.5\times10^6$.
The convection zone will shrink away from the
hot shell and back to a region where temperature is
less than $2.5\times10^6$ after the first $\sim$ 20 steps.
Thus, we expect some new born \iso{7}{Li} to survive.
The final abundance of \iso{7}{Li} depends on the creation and destruction processes in the hot shell.
In our merger models, there is no new carbon carried to the surface because
the convection zone does not reach the region of
$3\alpha$ burning which requires $T>1.2\times10^8$\citep{Suda2011}.

As in the  HeWD+HeWD merger, the temperature of hot shells depends on
the masses of the helium white dwarfs \citep{Zhang12a,zhu2013,Dan2013}.
Thus, a more massive HeWD will form a  hotter and broader
helium burning shell during the merger process.
Because of the higher temperature, less \iso{7}{Li} will survive,
but carbon from $3\alpha$ burning will be elevated to the surface by convection.

For the intermediate mass HeWDs, the surface convection zone does not contact the
helium burning region, but still reaches a layer where it is hot enough to destroy
almost all fresh \iso{7}{Li}.
Figure~\ref{5} shows that the remnants of the merger are divided into three groups,
i.e., Li-rich giants, C-rich giants and normal stars, which depend on
the HeWD progenitor masses.
Hence, we will discuss the mass ranges for these different cases separately.

\subsection{Li-rich red clump stars}

\begin{figure}
\includegraphics [angle=0,scale=0.5]{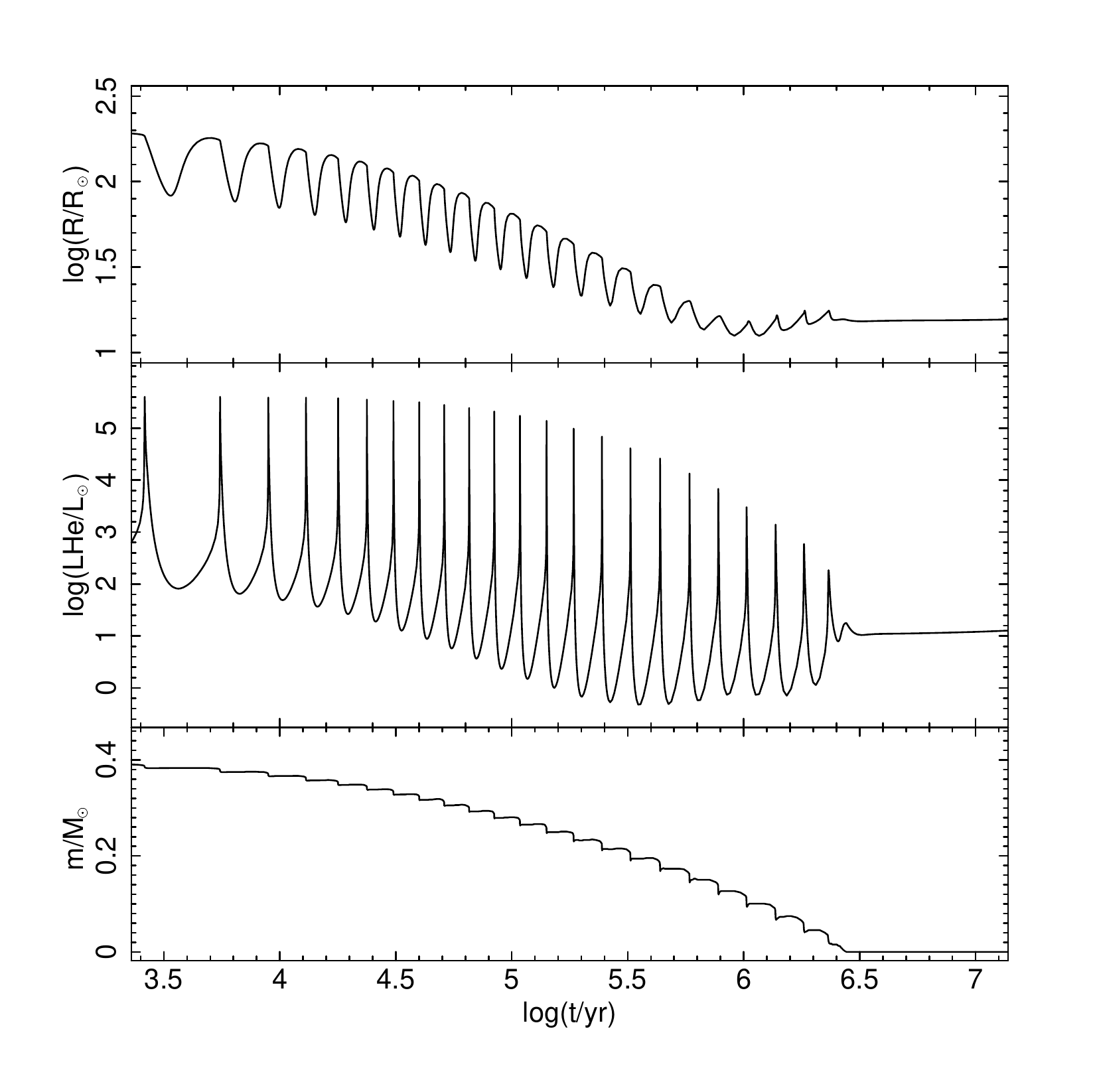}
\caption{The post-merger evolution of a 0.40\Msolar HeWD+0.15\Msolar helium core RGB
with a final mass of 1.10\Msolar before core burning. Top panel:
the evolution of radius. Middle panel: energy generated from the 3$\alpha$ reaction.
Bottom panel: the location of helium burning flame front.}
\label{6}
\end{figure}

\begin{figure}
\includegraphics [angle=0,scale=0.5]{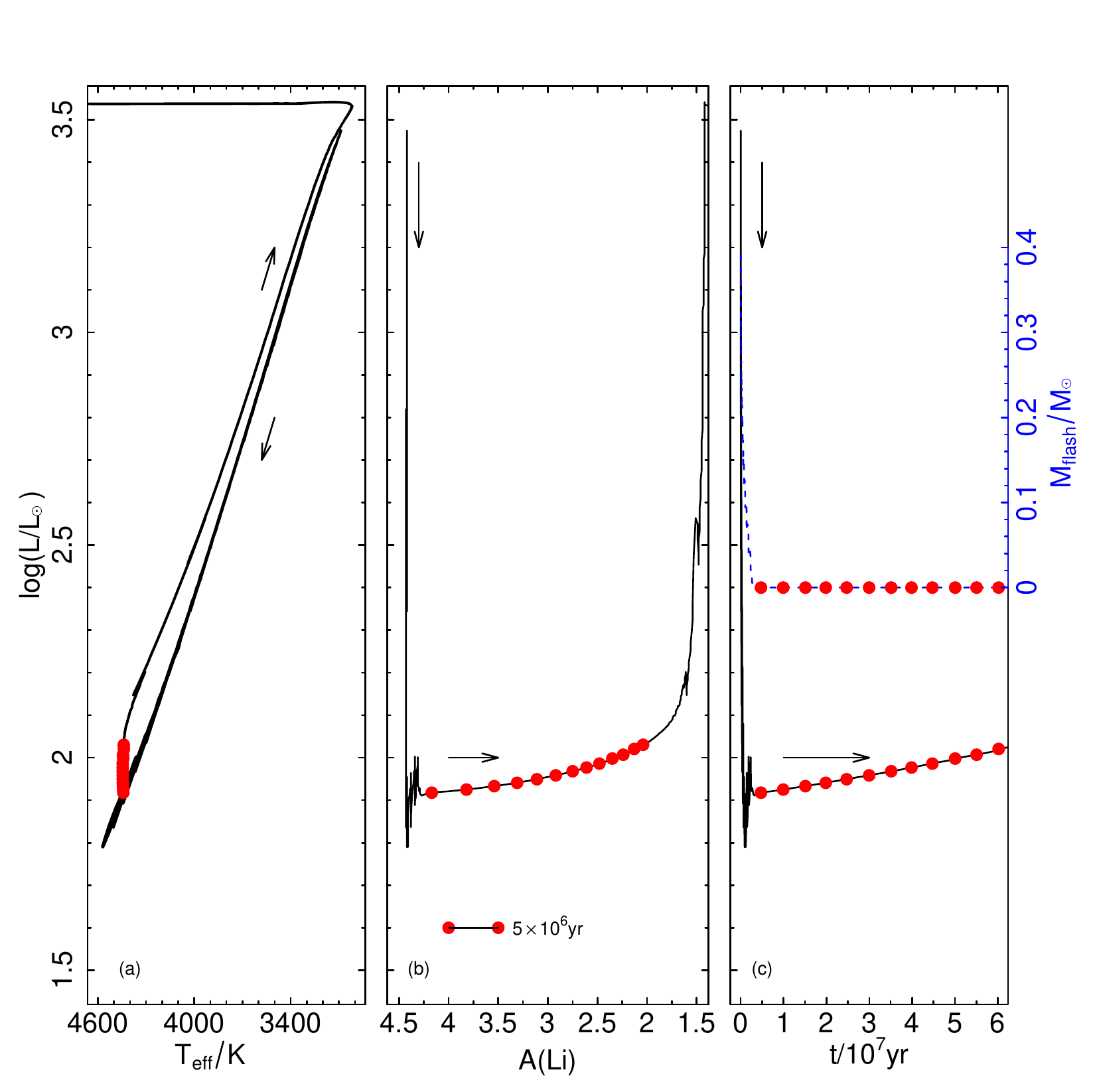}
\caption{Evolutionary tracks of the same model as in Fig.~\ref{6}. Left panel:
evolution in the HR-diagram. Middle panel: the evolution of surface abundance of lithium.
Right panel: the evolution of luminosity and helium flash location (dotted line, right-hand axis)
as a function of time. Red dots are separated in time by $5 \times 10^6$ yr.  }
\label{7}
\end{figure}

As discussed above, only the post-merger models formed from less massive HeWDs
can produce models with Li-rich surfaces, i.e., HeWD models with masses of
0.35,0.375 and $0.40\Msolar$ in our calculations.
As shown in Fig.~\ref{6}, the radius of star decreases during the inwards helium flash
before core burning. The inwards helium-core shell burning takes about 3 Myr.
Then the helium burning flame turns to outwards and stars in red clump stages.
The abundance of lithium decreased during core burning phase,
which takes tens of million years (see Fig.~\ref{7}). Once core helium burning ceases,
stars evolve to the AGB phase.

Considering that our post-merger models
spend most of their Li-rich stage in the helium core burning phase,
we expect to observe the majority as red clump stars.
Most of the Li-rich giants are close to the location of the
red clump on either the HR diagram or $T_{\rm eff}-\rm log g$ plane
(e.g., \citealt{Kumar2011,Adamow2014,Casey2016,Deepak2019,Zhou2019,Singh2019b}).
However, it is difficult to distinguish whether an observed giant is a red clump giant or is a degenerate-core giant ascending the RGB  for the first time (a red bump giant) from their
external parameters alone, i.e., from their luminosity, gravity, and temperature.
The best way to distinguish between these two stages is by asteroseismology \citep{Bedding2011}.
About 30 Li-rich stars been examined by this means (e.g., \citealt{Jofre2015,Carlberg2015,Kumar2018,Singh2019a,Singh2019b}).
Having found that model HeWD+RGB merger remnants can
become Li-rich red giants, we compare their
properties  with observed examples in more detail.

\subsubsection{Luminosity,  effective temperature and surface gravity}

\begin{figure}
\includegraphics [angle=0,scale=0.5]{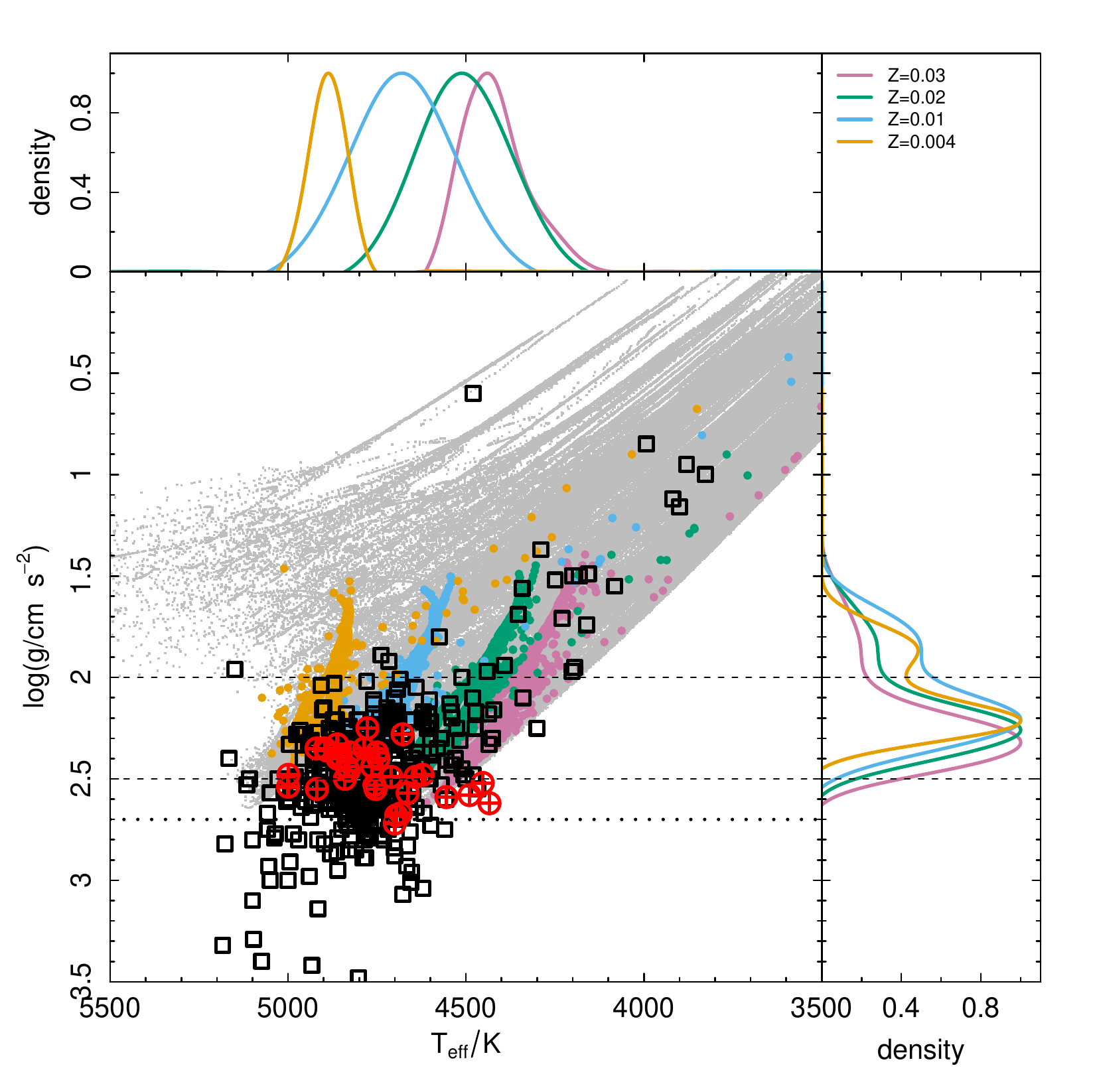}
\caption{Lithium-rich giants  in the effective temperature--surface gravity plane.
The grey dots indicate all the theoretical HeWD+RGB merger tracks computed which produce Li-rich giants.
The colored dots show the possible locations of Li-rich models from BSPS,
with different colours for different metallicity $Z$ as labelled in the key (top right).
The squares represent observed Li-rich giants without asteroseismology.
The 30 known Li-rich red clump stars are shown as circles with crosses.
 The dashed lines delimit the most likely range (in $\log g$) for Li-rich giants from BSPS.
The dotted line marks the maximum gravity of our Li-rich models.
 In the top-left and lower-right panels, curves colour coded for metallicity represent the normalized number density distributions of the Li-rich BSPS models in $T_{\rm eff}$ and $\log g$ respectively.
}
\label{8}
\end{figure}

\begin{figure}
\includegraphics [angle=0,scale=0.5]{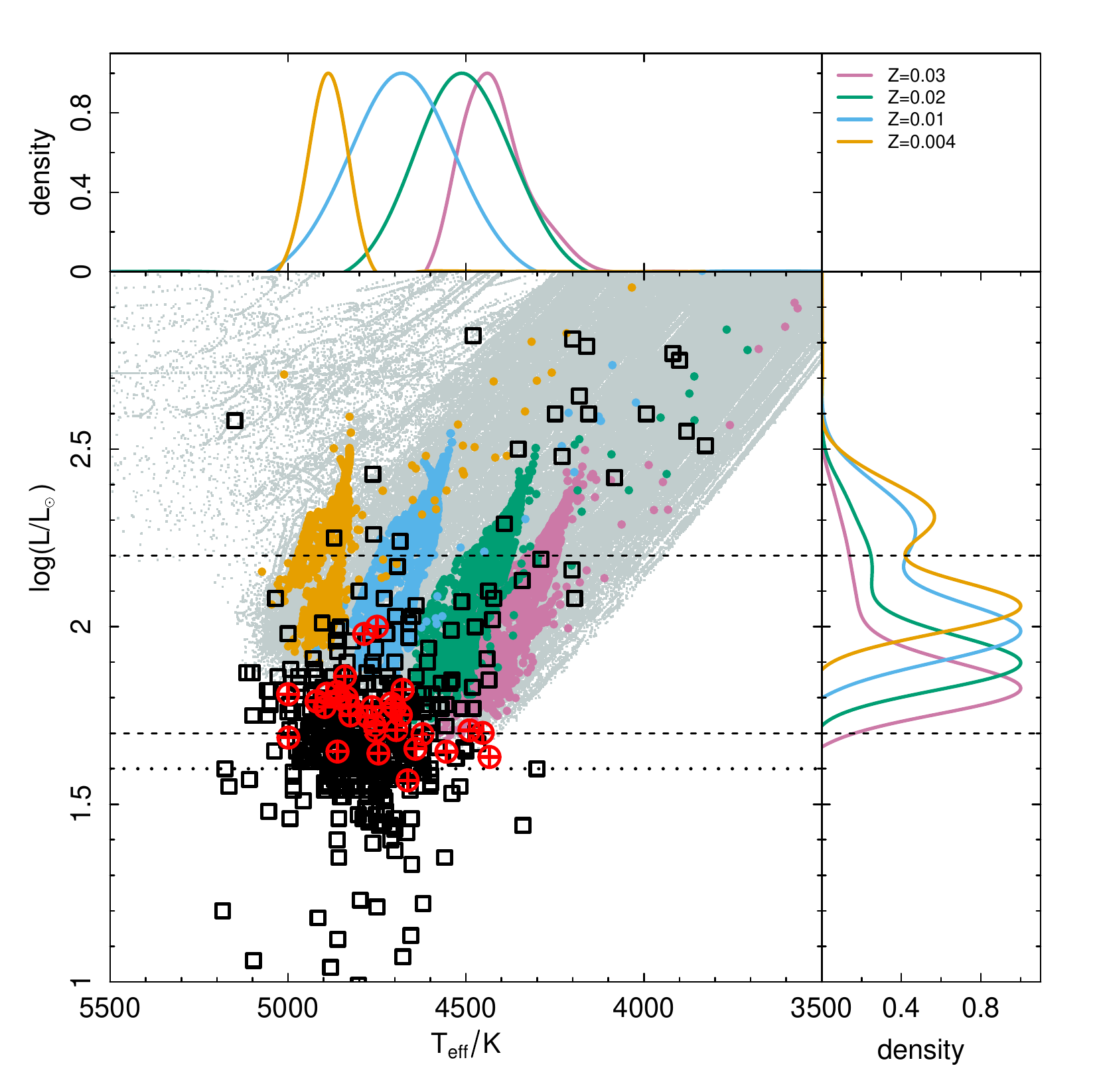}
\caption{As Fig.~\ref{8} for  $L - T_{\rm eff}$ plane (HR-diagram). }
\label{9}
\end{figure}

We compile a sample of 439 observed Li-rich giants for which parameters of luminosity,
surface effective temperature and gravity have been published
\citep{Kumar2011,Jofre2015,Carlberg2015,Kumar2018,Deepak2019,Zhou2019,Singh2019a,Singh2019b}.
Of these,  the evolution stage has been determined for 30.
All 30 are red clump stars \citep{Jofre2015,Carlberg2015,Kumar2018,Singh2019a,Singh2019b}.
In the $\log g - T_{\rm eff}$ plane, their locations
are similar to the  theoretical loci of Li-rich giants
formed through HeWD+RGB mergers (Fig.~\ref{8}).
The 30 known Li-rich red clump stars match the peak of the theoretical distribution very well.
A similar comparison can be made in the $L - T_{\rm eff}$ plane (HR-diagram: Fig.~\ref{9}).

\subsubsection{Surface abundance}

\begin{figure}
\includegraphics [angle=0,scale=0.5]{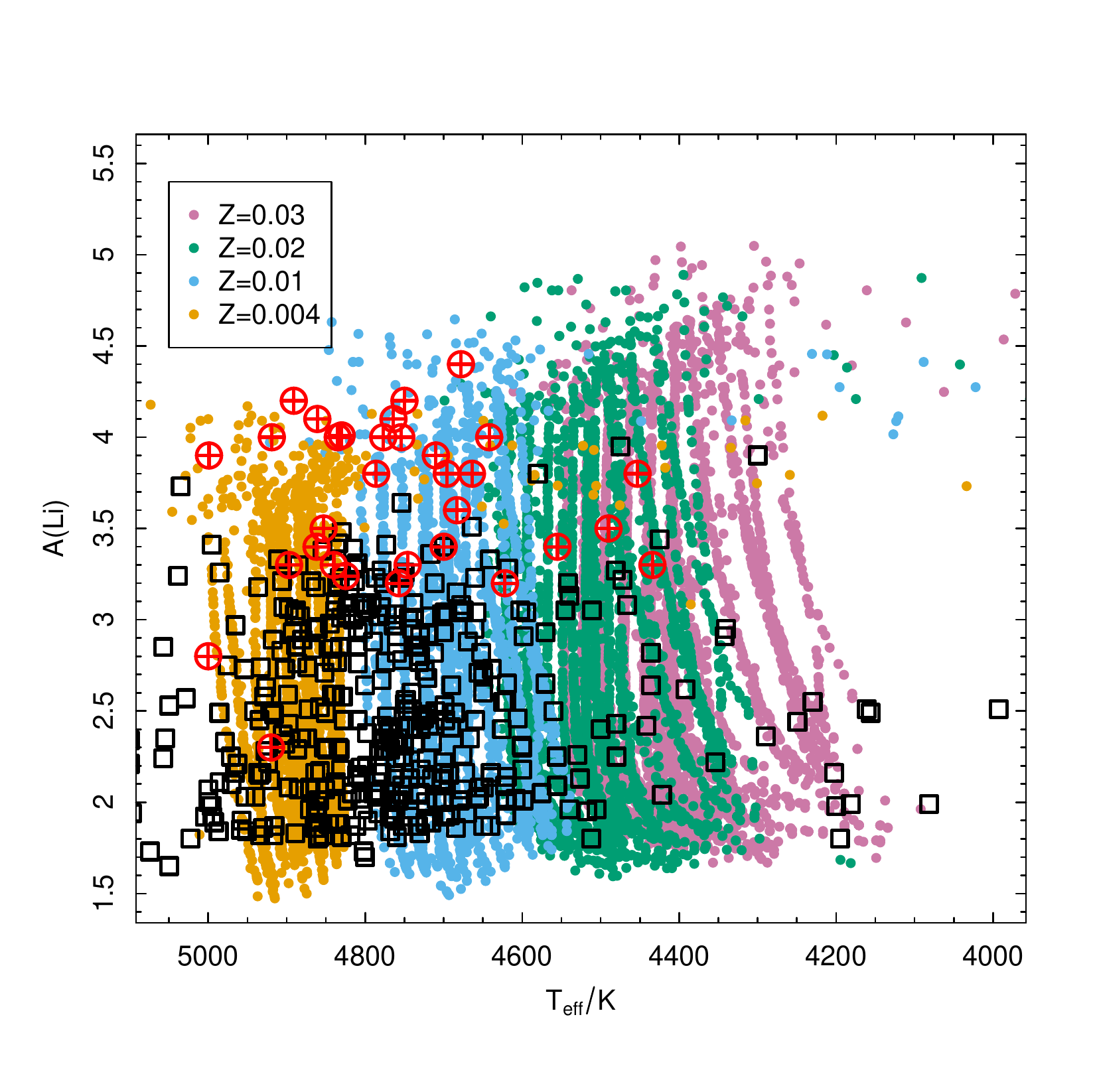}
\caption{Observed Li-rich giants in the effective temperature -- surface lithium (A(Li)) plane.
Symbols are as in Fig.~\ref{8}.}
\label{10}
\end{figure}

The most important feature to compare with
observation is the abundance of lithium.
Fig.~\ref{10} shows the lithium distribution of observation stars on A(Li)-T plane,
and almost all can represent by our models.

\begin{figure}
\includegraphics [angle=0,scale=0.5]{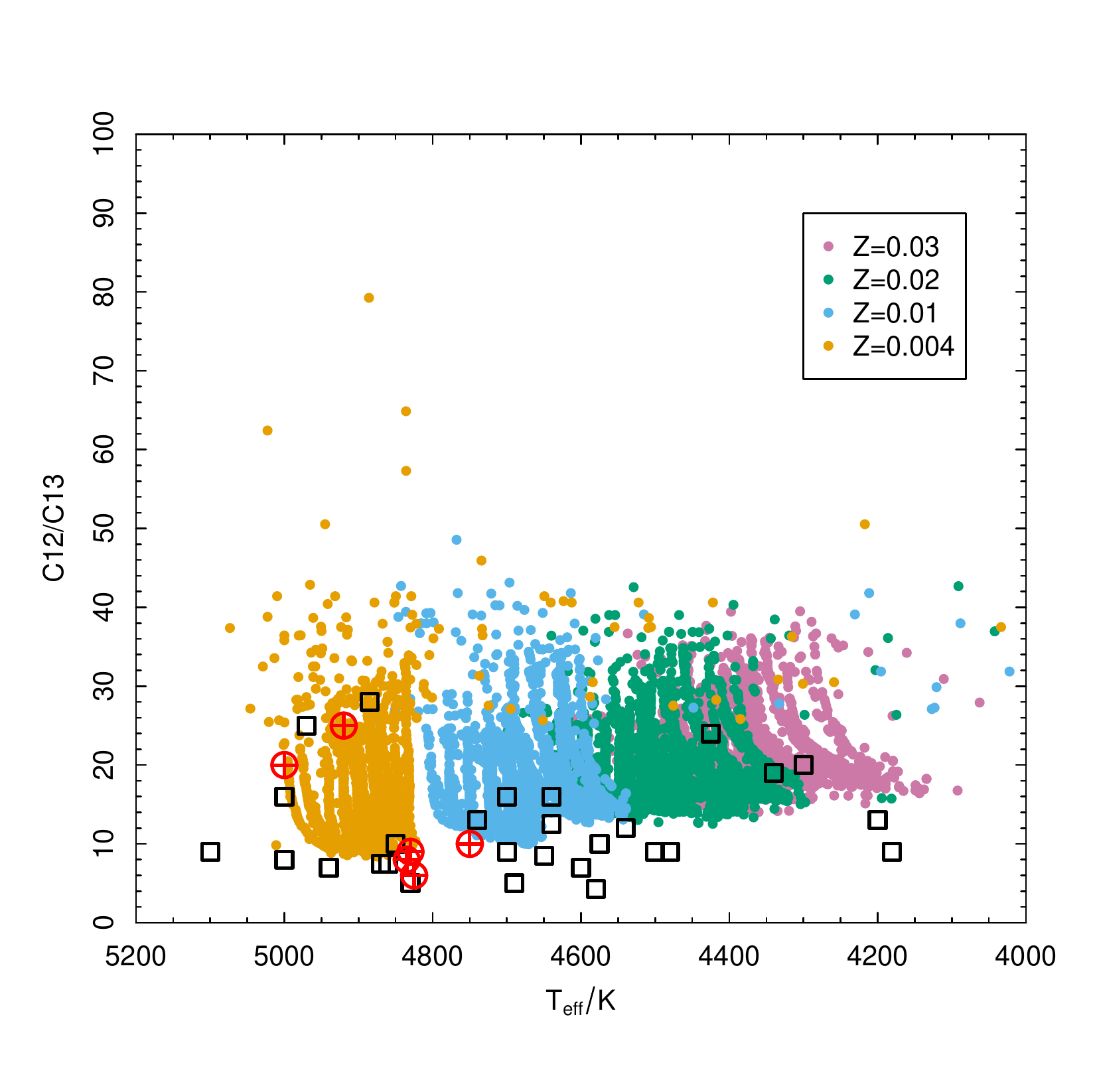}
\caption{Observed Li-rich giants in the effective temperature -- surface $^{12}\rm C/^{13}\rm C$ plane.
Symbols are as in Fig.~\ref{8}. }
\label{11}
\end{figure}

In addition to the abundance of the lithium, carbon isotope ratios are known for a few Li-rich giants, most of
which show a low  $^{12}\rm C/^{13}\rm C$ ratio ($<15$) \citep{Kumar2011}.
We compiled a sample of 39 Li-rich giants with a known  $^{12}\rm C/^{13}\rm C$ ratio to compare with our models (Fig.~\ref{11}).
Most of the observed ratios match the theoretical distribution of $^{12}\rm C/^{13}\rm C$ in Li-rich giant models.
The enrichment of \iso{13}{C} is due to the ${\rm ^{12}C}(p,\gamma){\rm ^{13}N}(\beta^{+},\nu){\rm ^{13}C}(p,\gamma){\rm ^{14}N}$
reaction followed by convective mixing (or dredge-up) to the surface during the merger.
The final fraction of \iso{13}{C} depends on the balance of the competition of \iso{13}{C} and \iso{14}{N} in the nuclear reactions.

\subsubsection{red clump status}

\begin{figure}
\includegraphics [angle=0,scale=0.5]{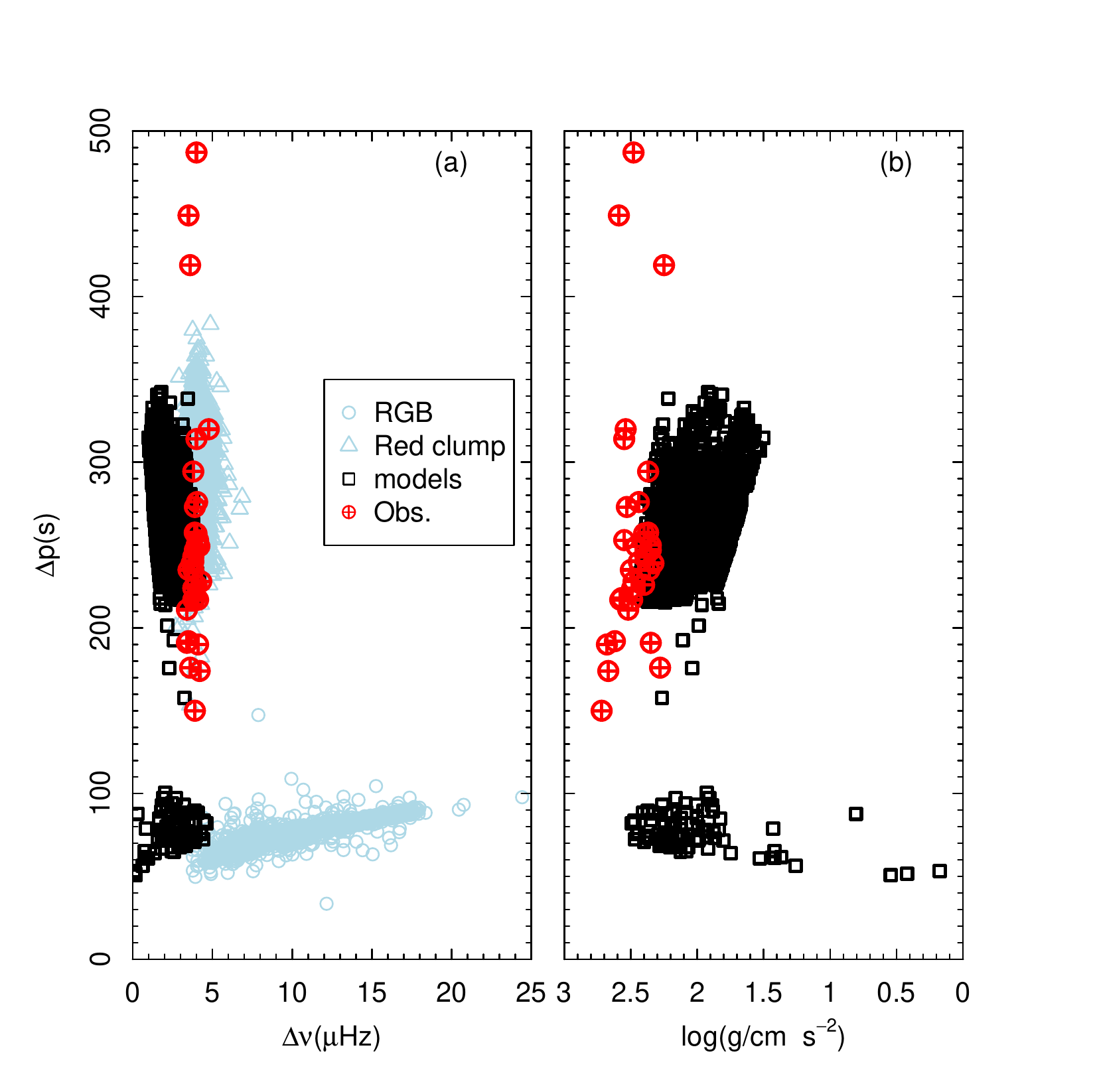}
\caption{Location of the sample stars and models on the seismic
diagram (left) and log g--$\Delta p$ plane (right). The red clump stars and
RGB stars are indicated by triangles and circles, which are from \citet{Vrard2016}.
The squares and circles with crosses indicate the
models and observed Li-rich giants, respectively. }
\label{12}
\end{figure}

As stated above, the post-merger models spend several million years with a
helium-burning shell moving inwards though a helium core.
Once the helium flash front reaches the center, actual core helium burning begins.
They then spend most of their lifetime as red giants in the red clump stage, i.e., tens of million years.
We would therefore expect most Li-rich giants to be red clump stars if they are formed by a merger.
As shown in Fig.~\ref{12}, the helium-core-burning stars (red clump stars) and hydrogen-shell-burning
stars (RGB stars) occupy different regions of the $\Delta \nu - \Delta p$ diagram\footnote{
$\Delta \nu$ and $\Delta p$ represent the large frequency and large period spacing which, in asymptotic theory, are characteristic of the spacing between successive radial orders of p- and g-mode oscillations of the same (non-radial) degree which, in turn, primarily reflect properties of the stellar envelope and core, respectively.}.
The red clump stars typically have $\Delta \nu/\mu{\rm Hz} \sim 5$ and  $200 < \Delta p/{\rm s} < 400$.
The RGB stars typcially have $\Delta p/{\rm s} \sim 100$ and $4 < \Delta \nu/\mu{\rm Hz} < 20$.
In other words, the red clump stars demonstrate a large range of core properties,
whilst the RGB stars demonstrate a large range of envelope properties.
All 30 Li-rich giants are red clump stars, and most match those theoretical models allowed by BSPS.
The same stars and models are shown on the  $\log g-\Delta p$ plane (also Fig.~\ref{12}) and also match well.
We calculate the asteroseismic quantities from simple scaling relations
(e.g.,\citealt{Ulrich1986,Brown1991,Kjeldsen1995}) using MESA output, and not from a full pulsation analysis  \citep[][e.g. GYRE]{Townsend2013}.
The former are used successfully for asteroseismic analyses of many classes of pulsating star.

Fig.~\ref{12} shows that there are a few post-merger models
in the RGB dominated region of the seismic diagram.
Such models are in the helium-core shell burning phase, where the
helium-burning shell is moving inwards into an electron-degenerate helium core.
Their density structure is thus similar to that of RGB stars
where the hydrogen-burning shell around an election-degenerate helium core is moving outwards.

\subsubsection{Masses}

\begin{figure}
\includegraphics [angle=0,scale=0.5]{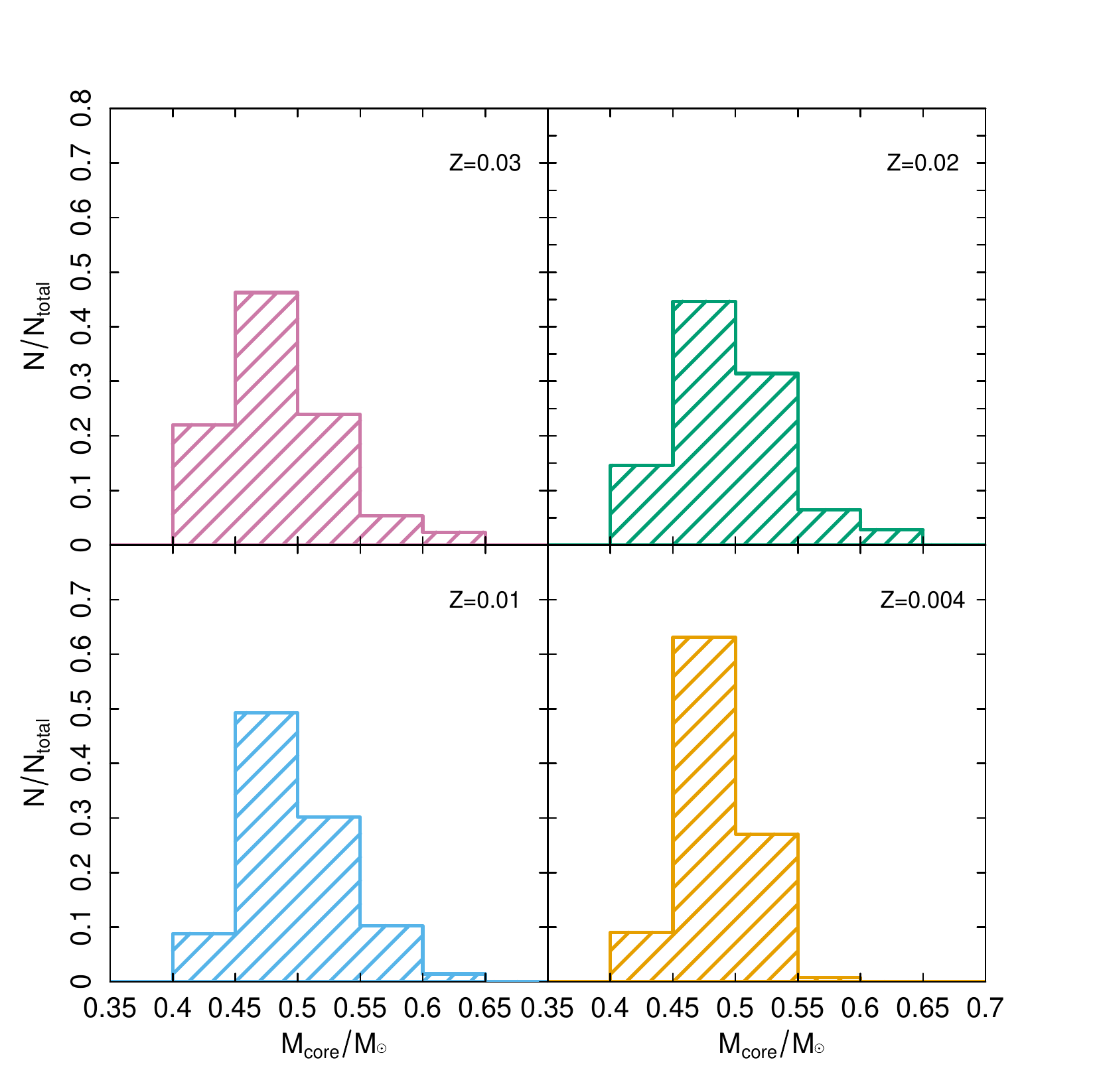}
\caption{The core mass distribution of Li-rich giant models at different metallicities. }
\label{13}
\end{figure}

\begin{figure}
\includegraphics [angle=0,scale=0.5]{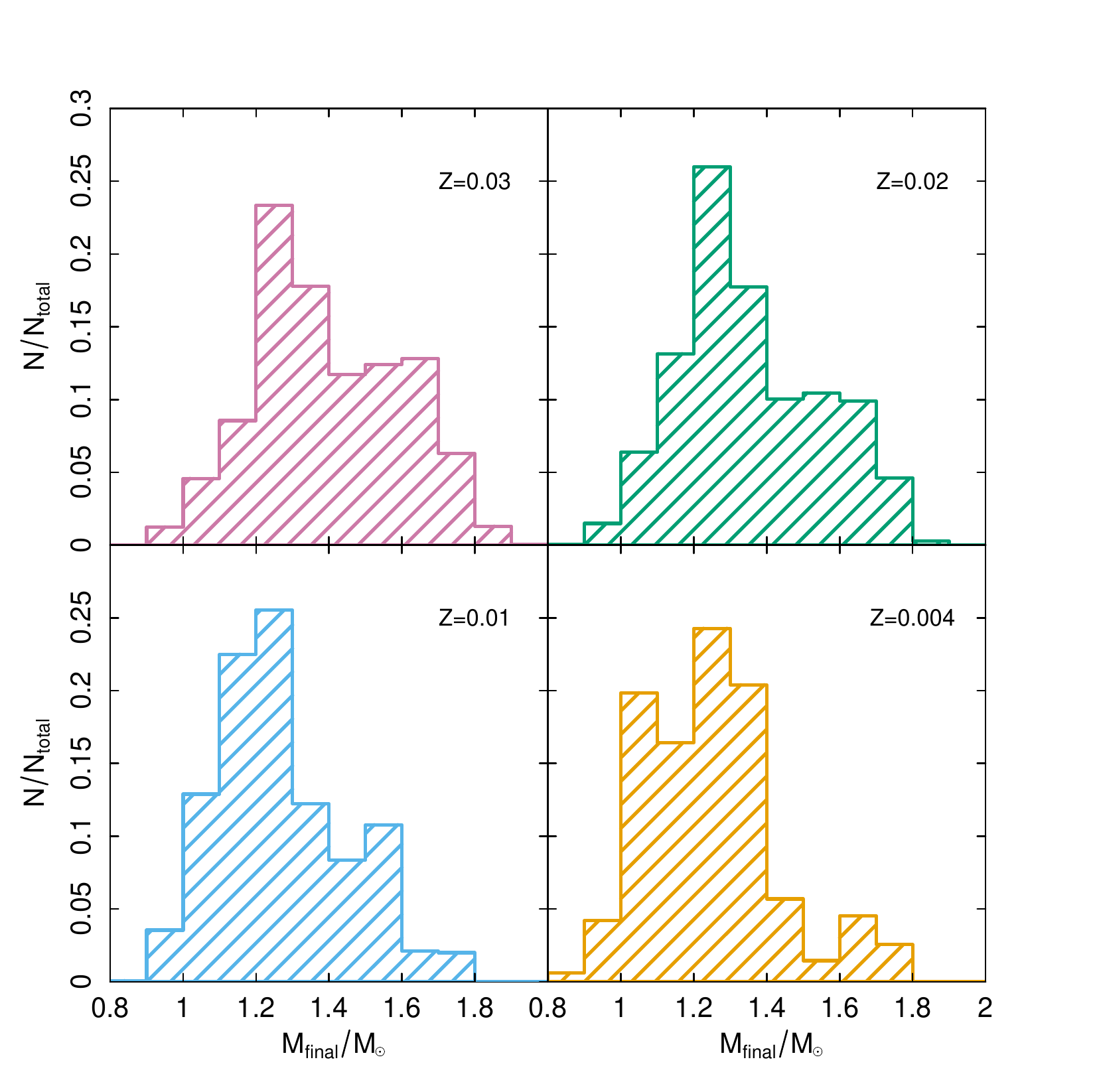}
\caption{The total mass distribution of Li-rich giant models at different metallicities. }
\label{14}
\end{figure}

Fig.~\ref{13} shows the distribution of the core masses of merger remnants.
The core masses are in the range $0.4-0.65\Msolar$, with a peak at $0.45-0.5\Msolar$.
Fig.~\ref{14} shows the final masses of Li-rich mergers, which range from 0.8 to 1.8$\Msolar$.
The peak of the mass distribution is $\approx1.2\Msolar$.

\begin{figure}
\includegraphics [angle=0,scale=0.5]{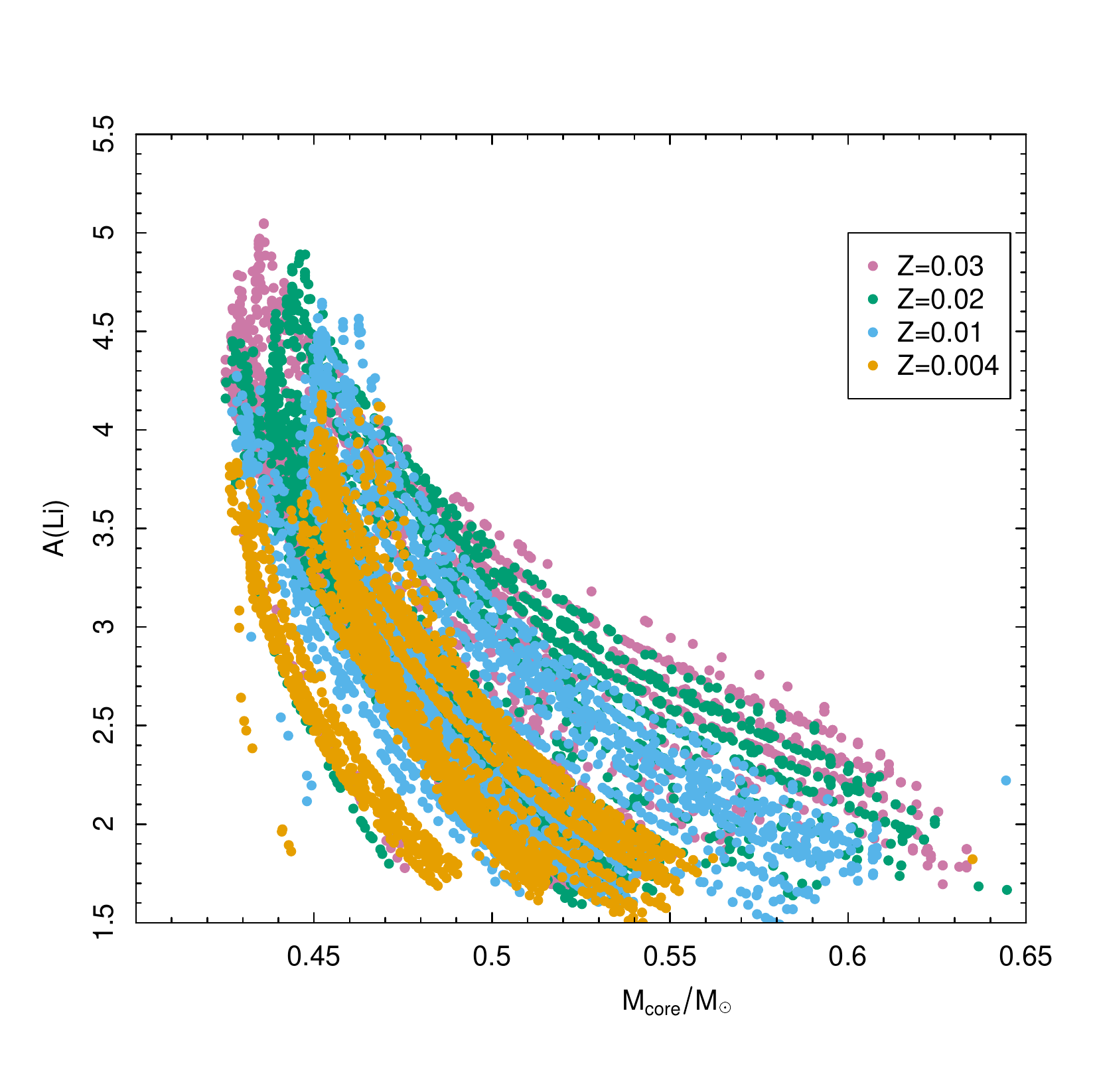}
\caption{The core mass--A(Li) plane. Symbols are as in Fig.~\ref{8}.}
\label{15}
\end{figure}

As discussed in section 2, the abundance of lithium in the post-merger models is very sensitive
to the mass of the pre-merger HeWDs and hence to the core masses of the remnants.
Fig.~\ref{15} shows how the surface abundance of lithium increases with decreasing core mass.
For stars with lithium $\rm A(Li)\ge 3.2$, the core masses are approximately in the range $0.44-0.51\Msolar$.

\subsubsection{Birth rate}

Aside from the question of how lithium is enriched by HeWD+RGB mergers,
a critical issue is how many such stars could be observed in the Galaxy.
To find the answer, we combine post-merger evolution models with the results of population synthesis
and a statistical analysis.
Our calculation is designed to find the properties of a model population
with an age of $14\,\rm{Gyr}$ and a constant star formation rate
of $5\,\Msolar\,\rm{yr}^{-1}$, which is designed to represent star-formation history of the Galaxy \citep{yungelson98}.

From a calculation of $10^7$ binary systems,
we find that  3931, 3233, 2707 and 3093 undergo HeWD+RGB mergers
and produce enriched lithium surfaces for metallicities $Z=0.03, 0.02, 0.01$ and 0.004, respectively.
Thus, at an age of  $13.7\,\rm{Gyr}$ with a constant star formation rate
of $5\,\Msolar\,\rm{yr}^{-1}$,
Li-rich giants form through HeWD+RGB mergers at a rate of 8.6, 7.0, 5.9
and $6.7 \times 10^{-4}\,{\rm yr}^{-1}$ for Galaxy disc at $Z=0.03, 0.02, 0.01$ and 0.004, respectively.

\begin{figure}
\includegraphics [angle=0,scale=0.5]{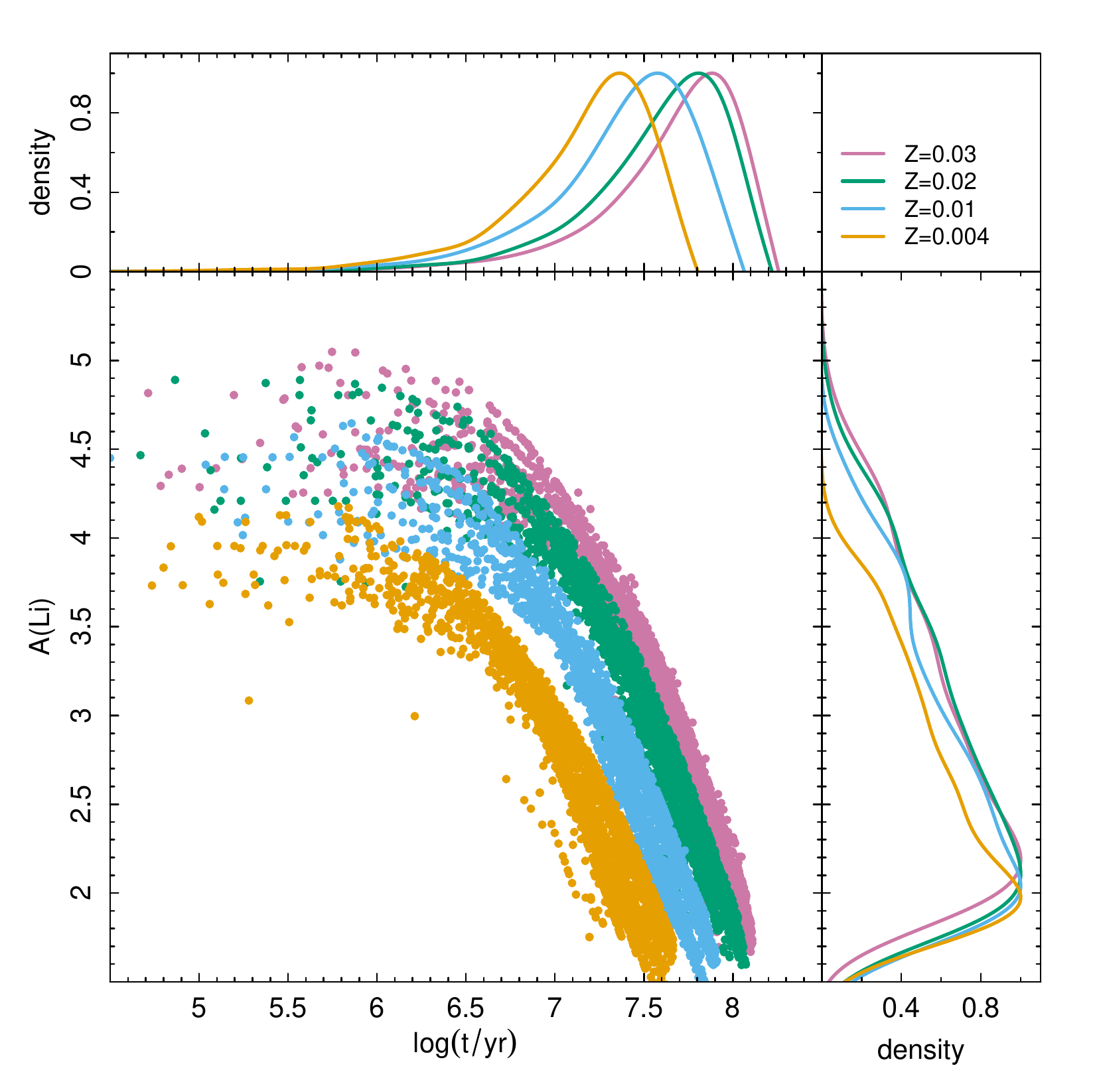}
\caption{Surface lithium abundances of Li-rich models as a function of age. Symbols are as in Fig.~\ref{8}.}
\label{16}
\end{figure}

Another factor to consider is the lifetime of Li-rich giants.
Lithium will decrease during evolution as surface lithium is dredged down and destroyed.
Fig.~\ref{16} shows the ages of Li-rich giants in our models since the post-merger stage.
The most enriched lithium giants only exist as such for a few millions of years but,
in total, the Li-rich giants defined as having $\rm A(Li)>1.5$ may live for up to $1.0 \times 10^8$ yrs.
Assuming a birth rate of $6-9 \times 10^{-4}\,{\rm yr}^{-1}$ and taking an effective Galactic disc volume of $7.5\times 10^{10} \rm pc^{3}$  \citep[e.g.]{YU2015}  ,
we expect the Galactic disc volume density of Li-rich giants to be $0.8-1.2\times 10^{-6} \rm pc^{-3}$,
assuming  that the majority are Li-rich red clump stars.
The apparent space density of red clump stars is about $1.14\times 10^{-4} \rm pc^{-3}$ in the disk of Galaxy \citep{Knapp2001}.
The predicted number of Li-rich red clump stars formed from mergers thus corresponds to $\sim 1\%$ of the total red clump stars.
This suggests that HeWD+RGB mergers contribute to most of the observed Li-rich red clump population.
However, some Li-rich giants may form from other channels, in particular the Li-rich red-bump RGB stars.

\subsection{Early-R carbon stars}

\citet{Zhang2013} argued that the early-R carbon stars could form from HeWD+RGB mergers.
In the current merger calculations we can reproduce the features of most early-R carbon stars.
As with Li-rich stars from the HeWD+RGB merger channel, the models that resemble eary-R stars are all in the red clump (core helium-burning) phase,
but with the difference that these models are formed from more massive helium white dwarfs.
Moreover, we only obtain carbon-rich models with metallicities $Z=0.03,0.02$ and 0.01, and {\it not} with $Z=0.004$.

\begin{figure}
\includegraphics [angle=0,scale=0.5]{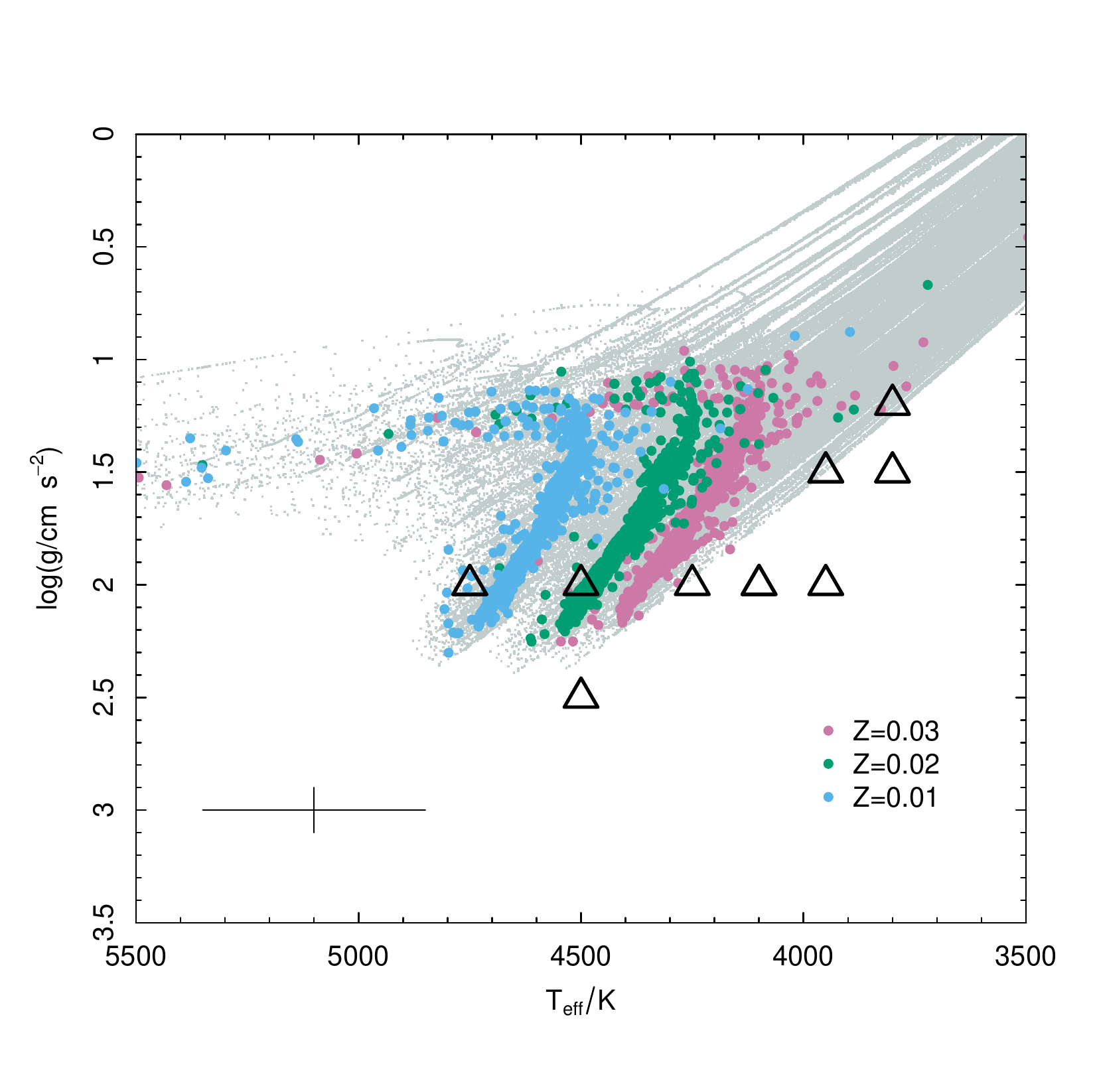}
\caption{Early-R stars in the effective temperature--surface gravity plane.
The grey dots indicate the theoretical distribution of C-rich models formed through HeWD+RGB mergers.
The colored dots show the possible location of stars from binary population synthesis, with different colours for different metallicities.
The triangles represent observed early-R stars from \citet{Zamora2009}.
The cross indicates the average error on the R-star observations. }
\label{17}
\end{figure}

Having found that model HeWD+RGB merger remnants can become early R stars,
we compare their properties in more detail to observed examples of such stars.
We have a sample of 12 early-R stars from \citet{Zamora2009}. We compare these stars with the theoretical distribution of atmospheric parameters for models formed through this merger channel .
The observed stars are satisfactorily close to the theoretical distribution in the
effective temperature--surface gravity plane (Fig.~\ref{17}).

\subsubsection{General properties}

\begin{figure}
\includegraphics [angle=0,scale=0.5]{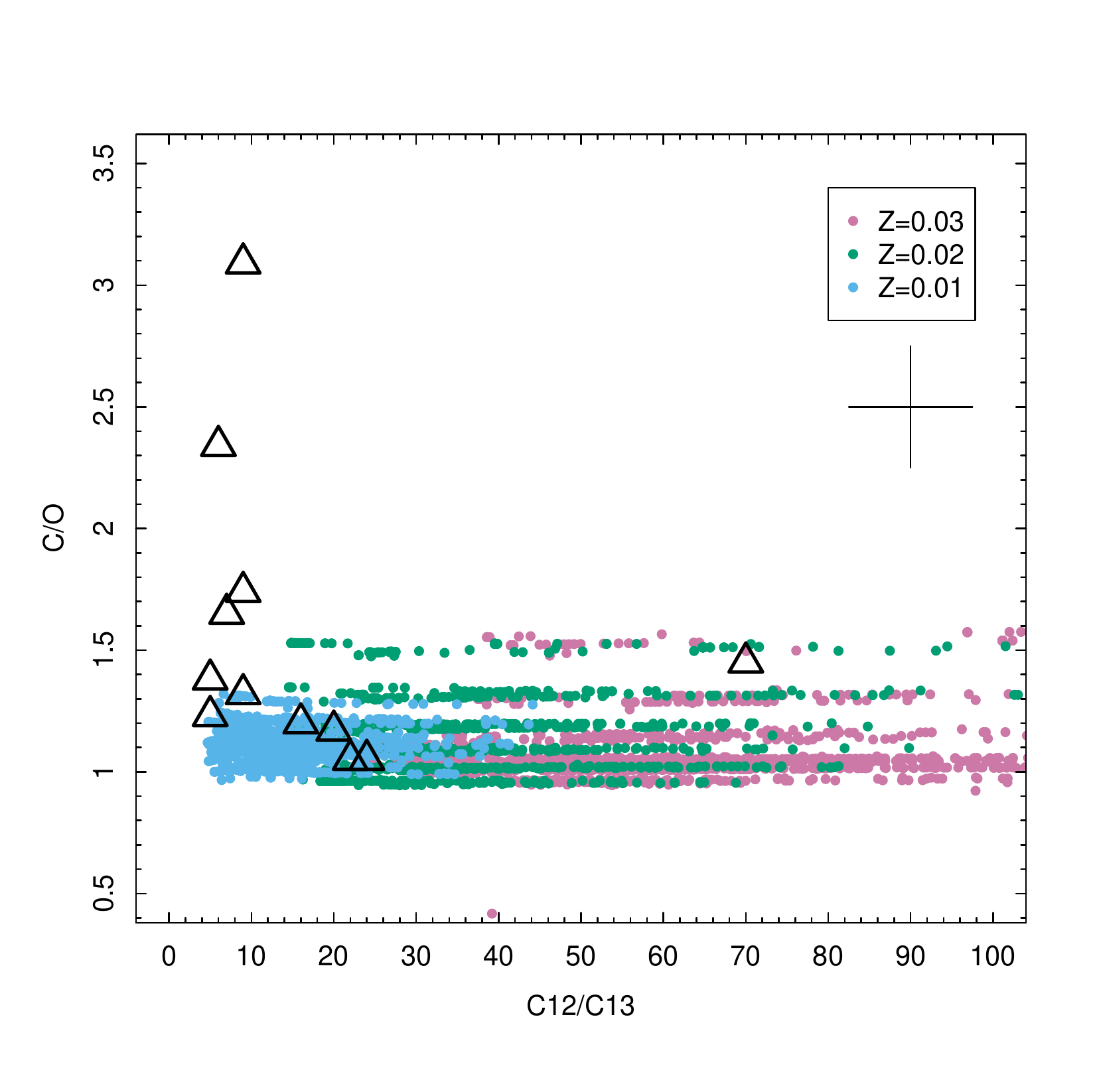}
\caption{As Fig.~\ref{17} for $\rm C/O$  and $\rm ^{12}C/^{13}C$ ratios.  }
\label{18}
\end{figure}

\begin{figure}
\includegraphics [angle=0,scale=0.5]{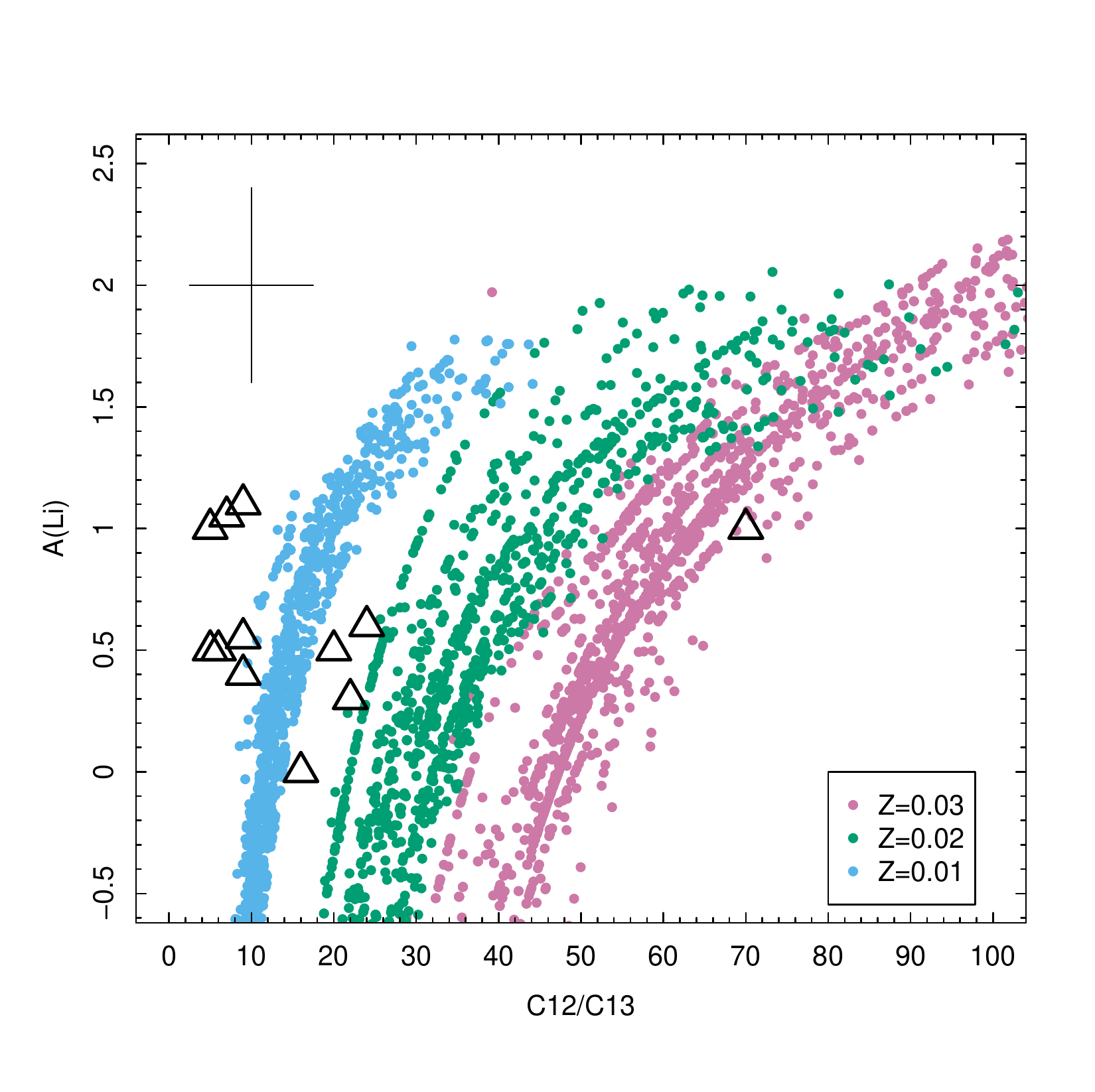}
\caption{As Fig.~\ref{17}  for A(Li) and $\rm ^{12}C/^{13}C$. }
\label{19}
\end{figure}

Fig.~\ref{18} compares the $\rm C/O$ and $\rm ^{12}C/^{13}C$ ratios for
the early-R star observations with ratios calculated in the models.
The theoretical models overlap the distribution of early-R stars with ${\rm C/O} \approx 1-2$.
Some of the stars with large ${\rm C/O}$ ratio may form from other channels.
The observed lithium abundances are shown in Fig.~\ref{19} and are consistent with the range obtained in the merger models.
The observed $\rm ^{12}C/^{13}C$ ratios are consistent with models having $Z\sim 0.01$.
Of the 12 stars, 8 have [M/H]<=-0.3 ($Z\sim 0.01$) and the rest are -0.26,
-0.1,-0.09,-0.03 \citep{Zamora2009}. Thus a metal-poor origin appears probable.

\subsubsection{Birth rate}

As in the study of Li-rich giants, we  want to know how many early-R stars formed
from merger we can expect to observe in the Galaxy.
Form the BSPS simulation of  $10^7$ binary systems,
we find 2051, 1262 and 1742 systems which undergo HeWD+RGB mergers
to produce enriched carbon stars, with $Z=0.03, 0.02$ and 0.01, respectively.
Thus, at $13.7\,\rm{Gyr}$, with a constant star formation rate
of $5\,\Msolar\,\rm{yr}^{-1}$,  C-rich giants are formed through HeWD+RGB mergers at a rate of 4.4, 2.6
and $3.8 \times 10^{-4}\,{\rm yr}^{-1}$ for Galaxy disc at $Z=0.03, 0.02$ and 0.01, respectively.

The lifetime for C-rich giants is about $10^8$ yrs.
Combining with the birth rate and the volume of the disk of Galaxy,
we may expect the volume density of early-R stars to $3.5-5.9\times 10^{-7} \rm pc^{-3}$.
The observed volume density of early-R stars is about $0.45-1.5\times 10^{-7} \rm pc^{-3}$ \citep{Knapp2001}.
These are small number statistics, but they do indicate that HeWD+RGB mergers can contribute for the early-R stars population.
Other channels may account for the most carbon-rich early-R stars.

\section{Conclusion and discussion}
\label{s_conclusion}

The merger of  a helium white dwarf (HeWD) with a red giant branch (RGB) star in a common envelope
provides a possible model for the origin of lithium and carbon-rich stars.
By adopting a grid of approximate initial conditions for such mergers informed by previous SPH simulations,
we have made one-dimensional {\sc mesa} calculations of the post-merger evolution of a wide range of possible systems.
Using binary star population synthesis (BSPS) we have computed the frequency and distribution of HeWD + RGB binary-star systems that are likely to merge.

Analysis of the post-merger evolution models show that the final surface abundances depend on the masses of the progenitor HeWDs.
The Li-rich giants form from low-mass HeWDs, i.e., $\rm 0.35 \Msolar \le M_{WD}\le 0.40\Msolar$;
the C-rich giants form from low-mass HeWDs, i.e., $\rm 0.45 \Msolar \le M_{WD}\le 0.475\Msolar$.
The pathway to either these categories is related to the temperature of the helium core burning shell
and the balance of nuclear products during initial shell burning.

Analysis of the distribution of post-mergers systems, as given by BSPS and {\sc mesa} evolution tracks,
shows that predictions for HeWD+RGB mergers  are consistent with observations of most Li-rich giants in terms of surface effective temperature ($T_{\rm eff}$), surface gravity ($log g$), surface luminosity ($log L$), surface abundance (A(Li) and $^{12}\rm C/^{13}\rm C$),
and the Galaxy disc space density  of $0.8-1.2\times 10^{-6} \rm pc^{-3}$.
They are also consistent with the properties of some early R carbon stars, in terms of $\rm T_{eff}$,$\rm log g$,
$\rm C/O $, A(Li), $^{12}\rm C/^{13}\rm C$, and a space density of $3.5-5.9\times 10^{-7} \rm pc^{-3}$.

According to the models, the post-merger objects are helium core burning stars, which means that they are all in the red clump stage of evolution, i.e. they are red giants with non-degenerate helium cores.
However, only a few Li-rich giants have had their precise evolutionary stage determined by asteroseismology.
Those stars which have been analysed in this way correspond well with our results.
Hence, we strongly argue that the Li-rich red clump stars and the majority of early-R carbon stars
most likely form following the merger of a HeWD with an RGB star.

A few problems still need to be addressed:\\
i)  post-merger evolution calculations depend on evolution during the common envelope phase
which is determined by the progenitors of the merger,
the amount of mass loss during the common envelope phase and the structure of the common envelope.
However, our knowledge of common envelope phase in binary-star evolution is still poor.
Future theoretical studies and simulations will improve this deficiency.
Also, as the products of post common-envelope mergers, further observations of Li-rich red clump stars
will provide additional data for studies of the common-envelope phase.

ii) Some of the observed Li-rich giants have measured rotation velocities,
but we have not calculated rotation speeds for post-merger models immediately following common envelope ejection.
Our model will be improved with more knowledge of angular momentum transport
during the merger.

iii) The HeWD+RGB merger model provides a successful explanation for  Li-rich red clump stars,
but does not explain Li-rich RGB stars, i.e. stars on their first ascent of the giant branch. Such stars
will require further explanation in the future.

iv) The most carbon-rich early-R  stars (with C/O$\gg1.5$) cannot be explained by our models.

We therefore need more observations of Li-rich and C-rich giants
to help identify and diagnose the evolution channels.
We also require further numerical simulations for HeWD+RGB mergers, in particular to study the dynamics of the common envelope and the merging of the HeWD with the RGB core.

\acknowledgments
We would like to thank the helpful suggestions and comments from the referee that
improved the manuscript. XZ thanks Hongliang Yan, Haining Li, and Yerra Bharat Kumar
for helpful conversations.
This work is supported by the grants 11703001, 10933002 and 11273007 from the National Natural Science Foundation of China,
the Joint Research Fund in Astronomy (U1631236) under cooperative agreement between
the National Natural Science Foundation of China (NSFC) and Chinese Academy of Sciences (CAS),
and the Fundamental Research Funds for the Central Universities.
Armagh Observatory and Planetarium is supported by a grant from the Northern Ireland Department for Communities.

\bibliographystyle{apj} % (uses file "abbrvna.bst")
\bibliography{mybib} % expects file "mybib.bib"

\appendix
\section{\textsc{mesa} inlist}\label{sec:mesa_inlist}
To evolve merger remnants with \textsc{mesa} the parameters that differ from the defaults are as follows:
{\small
\begin{verbatim}

&star_job
   change_net = .true.
   new_net_name = 'agbnew.net'

   set_rates_preference = .true.
   new_rates_preference = +2

   kappa_file_prefix = 'a09'
   kappa_lowT_prefix = 'lowT_fa05_a09p'
   kappa_CO_prefix = 'a09_co'

   initial_zfracs = 6 ! AGSS09_zfracs

/

&controls
   use_Type2_opacities = .true.
   kap_Type2_full_off_X = 1d-3
   kap_Type2_full_on_X = 1d-6
   Zbase = 0.02d0

   mixing_length_alpha = 1.82

   use_Ledoux_criterion = .true.
   alpha_semiconvection = 0.1
   thermohaline_coeff = 666.0

   which_atm_option = 'simple_photosphere'

   cool_wind_RGB_scheme = 'Reimers'
   Reimers_scaling_factor = 0.5
   cool_wind_AGB_scheme = 'Blocker'
   Blocker_scaling_factor = 0.5
   RGB_to_AGB_wind_switch = 1d-4

   varcontrol_target = 1d-3
   mesh_delta_coeff = 2

   do_element_diffusion = .true.
   diffusion_dt_limit = 3.15e13
   diffusion_min_T_at_surface = 1d3

/
\end{verbatim}
}

\end{document}